\documentclass[12pt,a4paper]{article}  
\usepackage[utf8]{inputenc}
\usepackage[T1]{fontenc}
\usepackage[colorlinks,citecolor=blue,urlcolor=blue,linkcolor=blue]{hyperref}
\usepackage{graphicx}
\usepackage{dcolumn}
\usepackage{bm}
\usepackage{amsmath}
\usepackage{amsfonts}

\usepackage{titlesec}
\titlespacing*{\section}{0pt}{2ex}{2ex}
\titlespacing*{\subsection}{0pt}{2ex}{2ex} 
\titlespacing*{\subsubsection}{0pt}{2ex}{2ex}

\titleformat*{\section}{\large\bfseries}
\titleformat*{\subsection}{\large\bfseries}
\titleformat*{\subsubsection}{\large\bfseries}
\titleformat*{\paragraph}{\large\bfseries}
\titleformat*{\subparagraph}{\large\bfseries}

\usepackage{amssymb}
\usepackage{mathrsfs}
\usepackage{braket}
\usepackage{bbm}
\usepackage{xcolor}
\definecolor{olive}{rgb}{0.3, 0.4, .1}
\definecolor{fore}{RGB}{249,242,215}
\definecolor{back}{RGB}{51,51,51}
\definecolor{title}{RGB}{255,0,90}
\definecolor{blackViolet}{RGB}{138,43,226}
\definecolor{dgreen}{rgb}{0.,0.6,0.}
\definecolor{gold}{rgb}{1.,0.84,0.}
\definecolor{JungleGreen}{cmyk}{0.99,0,0.52,0}
\definecolor{blackGreen}{cmyk}{0.85,0,0.33,0}
\definecolor{RawSienna}{cmyk}{0,0.72,1,0.45}
\definecolor{Magenta}{cmyk}{0,1,0,0}
\definecolor{wood}{RGB}{139,115,85}
\definecolor{dorange}{RGB}{255,127,0}
\definecolor{dolive}{RGB}{85,107,47}
\definecolor{drg}{RGB}{255,165,0}

\DeclareMathAlphabet{\mathpzc}{OT1}{pzc}{m}{it}

\usepackage{multirow}

\usepackage{tikz}
\usepackage{colortbl}
\usepackage{multimedia}
\usepackage{xcolor}

\usepackage{graphicx}
\usepackage{ragged2e}
\usepackage{array}
\usepackage{rotating}
\usepackage{longtable}
\usepackage{arydshln}
\usepackage{multirow}

\usepackage{simplewick}
\usepackage{dirtytalk}

\usepackage{mathtools}

\newcommand\tvecpre{{\mathpalette\raiseT\top}}
\newcommand\raiseT[2]{\raisebox{-0.4ex}{$#1#2$}}
\newcommand\tvec{\mkern-2mu^\tvecpre\!}
\newcommand{\tmat}{\mkern-2mu ^\top\!}

\usepackage{stmaryrd}
\usepackage{geometry}
\usepackage{cite}

\makeatletter
\newcommand\xLongLeftRightArrow[2][]{%
  \ext@arrow 0099{\LongLeftRightArrowfill@}{#1}{#2}}
\def\LongLeftRightArrowfill@{%
  \arrowfill@\Leftarrow\Relbar\Rightarrow}
\makeatother
\usepackage{amssymb} 

\usepackage{amsthm}

\newtheorem{lemma}{Lemma}[section]
\newtheorem{corollary}{Corollary}[section]
\newtheorem{proposition}{Proposition}[section]
\newtheorem{definition}{Definition}[section]

\newtheorem{example}{Example}[section]

\usepackage{dsfont}

\begin{document}


\title{\normalsize{\textbf{Auxiliary many-body wavefunctions for TDDFRT electronic excited states}} \\ \normalsize{\textbf{Consequences for the representation of molecular electronic transitions}}}%

\date{}

\maketitle
\vspace*{-2cm}

\noindent \begin{center}
\textit{by}\footnote{Affiliation of the authors is given in a dedicated section right before the bibliography.} Jérémy Morere, Enzo Monino, \textit{and} Thibaud Etienne
\end{center}

$\;$

\begin{abstract}
\noindent This contribution reports the study of a set of molecular electronic-structure reorganization representations related to light-induced electronic transitions, modeled in the framework of time-dependent density-functional response theory. More precisely, the work related in this paper deals with the consequences, for the electronic transitions natural-orbital characterization, that are inherent to the use of auxiliary many-body wavefunctions constructed \textit{a posteriori} and assigned to excited states — since time-dependent density-functional response theory does not provide excited state ansatze in its native formulation. Three types of such auxiliary many-body wavefunctions are studied, and the structure and spectral properties of the relevant matrices (the one-electron reduced difference and transition density matrices) is discussed and compared with the native equation-of-motion time-dependent density functional response theory picture of an electronic transition — we see for instance that within this framework the detachment and attachment density matrices can be derived without diagonalizing the one-body reduced difference density matrix. The common ``departure/arrival'' wavefunction-based representations of electronic transitions are also extensively discussed. \\ $\;$ 

\end{abstract}

\maketitle

\section{Introduction} 

Time-dependent density-functional response theory (TDDFRT \cite{casida_time-dependent_1995,casida_time-dependent_2009,dreuw_single-reference_2005,hirata_configuration_1999,ferre_density-functional_2016}) is probably the most used electronic excited-state calculation method for complex molecular systems. Such a method gives access to difference and transition properties — the first allows the comparison between the properties in the ground and in the excited state, while the second informs about the coupling between the states, and the response of the system to the light-induced perturbation of the system — but it does not come with an ansatz for the excited states that are computed, at least in its equation-of-motion (EOM) formulation. Knowledge of such an ansatz would however be very useful for analyzing the nature of excited states \cite{adamo_calculations_2013,luzanov_electron_2010,lipkowitz_excited_2009}, but also in the context of non-adiabatic excited-state dynamics \cite{ferre_density-functional_2016,barbatti_nonadiabatic_2011,wang_nactddft_2021,hu_nonadiabatic_2007,hu2010nonadiabatic}, for it allows the computation of quantities such as the non-adiabatic coupling between states \cite{li_first-order_2014,li_first_2014}. Therefore, multiple research groups have worked out expressions for approximate, auxiliary many-body wavefunctions (AMBW) that are constructed \textit{a posteriori}, i.e., based on the data and objects issued from the excited-state TDDFRT calculations. The ``existence'' of certain of these wavefunctions is sometimes conditioned by some constraints that will also be recalled in this paper.

The first type of these AMBW's \cite{ipatov_excited-state_2009} has been proposed in a famous paper where the author exposes a transposition of the time-dependent Hartree-Fock (TDHF \cite{mclachlan_time-dependent_1964,hirata_configuration_1999}) excited-state calculation procedure to the realm of electron densities, introducing the TDDFRT method \cite{casida_time-dependent_1995}. The construction of an AMBW in this paper — which has the same structure as a configuration interaction singles (CIS) wavefunction \cite{hirata_configuration_1999,mayer_using_2007,maurice_configuration_1995,foresman_toward_1992-1,surjan_natural_2007} — is physically motivated, and related to the expression of the TDDFRT transition moments, but as we will see in this report, the excited states constructed with this procedure are not mutually orthogonal, and an orthogonalization procedure makes the nature of the considered states dependent on the size and content of the set of auxiliary states that is initially considered. The ansatz itself has been extensively used in the context of excited-state dynamics simulations \cite{crespo-otero_recent_2018,plasser_surface_2014,tapavicza_trajectory_2007,tavernelli_non-adiabatic_2009,tavernelli_nonadiabatic_2009,tavernelli_nonadiabatic_2009-1,tavernelli_nonadiabatic_2010}. We will also discuss three other AMBW's of the same type — i.e., with a strict CIS--like structure.

The second type of AMBW was brought by Luzanov and Zhikol by solving an alternative TDDFRT response equation, yielding states that are normalized and mutually orthogonal \cite{luzanov_electron_2010}. The drawback of this procedure is that it produces coupled \textit{pairs} of ansatze rather than a single electronic excited-state wavefunction.

The third type of AMBW analyzed in this paper comes from the work of Subotnik and co-authors \cite{alguire_calculating_2015}. It is based on the Hellmann-Feynman theorem, includes Pulay terms associated with an atom-centered basis, as well as a correction ensuring translational invariance, and ensures a proper behavior near the crossing between two states and in the complete basis limit. However, as for the previous AMBW type, these ansatze come in pairs.

The analysis and comparison of the EOM-TDDFRT picture of an electronic transition with the AMBW properties will be performed in terms of some relevant one-body reduced density matrices (1--RDM's): the one-electron reduced unrelaxed difference density matrix (1--DDM) and the one-body reduced transition density matrix (1--TDM) \cite{plasser_new_2014,etienne_towards_2021,etienne_comprehensive_2021,luzanov_application_1976,dreuw_single-reference_2005}, and their spectral decomposition yielding the so-called natural difference (NDO's) and transition (NTO's) orbitals \cite{plasser_new_2014,etienne_towards_2021,dreuw_single-reference_2005,martin_natural_2003,mayer_using_2007,surjan_natural_2007}.

Since the principal AMBW dealt with in this report takes the form of a linear combination of Fock states that are singly-excited relatively to the Fermi vacuum, one section of this report will introduce a quick review of the basic properties of such ansatze: the CIS wavefunction itself will be introduced, and the properties of CIS electronic transitions will be given in terms of natural orbitals.

First type AMBW will then be introduced and its properties studied. The structure and natural-orbital properties of this model will be compared with the EOM--TDDFRT picture. Consequences of this representation for the one-body reduced transition density kernel will also be mentioned.

Finally, the AMBW introduced by Luzanov and Zhikol, and the one introduced by Subotnik and co-workers, will be exposed and analyzed. The detachment/attachment density matrices \cite{head-gordon_analysis_1995,dreuw_single-reference_2005} will be derived for the AMBW of Luzanov and Zhikol, while the AMBW of Subotnik and co-workers will be shown to elude the native TDDFRT de-excitation contributions when used for deriving the 1--TDM. The 1--DDM for that type of auxiliary ansatz has also been fully derived and contains 254 terms.

\section{Hypotheses and notations} \label{sec:hyp} 

References \cite{etienne_towards_2021,etienne_comprehensive_2021} are two strongly recommended prerequisite papers for this contribution. Bold Greek letters and bold, upper case letters from Latin alphabet will denote matrices with more than one column. For instance, while $(\textbf{x})_{ia}$ will denote the $ia$ component of vector \textbf{x} where $ia$ is assimilated to an individual integer, $(\tilde{\textbf{X}})_{i,a}$ will denote the $i\times a$ element of matrix $\tilde{\textbf{X}}$ and $(\textbf{A})_{ia,jb}$ will denote the $(ia)\times (jb)$ element of matrix $\textbf{A}$, where again $ia$ and $jb$ are assimilated to individual integers.

In this contribution we will consider $N$--electron systems, with $N$ a natural integer strictly superior to one. For any electronic state considered below, the $N$ electrons are described using an $L$--dimensional orthonormal basis of real-valued spinorbitals, with ($1 < N < L$), each spinorbital being denoted by an indexed $\varphi$ symbol and defined on $S_4 \coloneqq \mathbb{R}^3\times \llbracket 1, 2 \rrbracket $. Since we are working within the non-relativistic quantum-mechanical framework, every spinorbital $\varphi$ is the tensor product of a spatial part $\chi$ — belonging to the first-order Sobolev space $H^1 (\mathbb{R}^3,\mathbb{R})$ of real-valued square-integrable functions whose first derivative is also square-integrable — and a spin part $\sigma$, i.e., one of the two possible electron spin states, namely $\alpha$ and $\beta$, where $\alpha$ is the $(1 \; 0)\tmat$ column vector, and $\beta$ is the $(0 \;\, 1)\tmat$ column vector. One-body reduced state- and difference-density matrices in this contribution are all symmetric.

Any molecular electronic ground-state wavefunction considered in this manuscript is a single $N$--dimensional determinant state: When $L$ spinorbitals are being produced when solving the Kohn-Sham equations for optimizing the ground state wavefunction, only $N$ are ``occupied'' with occupation number equal to unity, while $(L-N)$ are ``unoccupied'' (or ``virtual'') with their occupation number equal to zero in the ground state. Accordingly, we define three integer intervals:
\begin{align*}
I \coloneqq \llbracket 1,N\rrbracket,\,
C \coloneqq \llbracket 1,L\rrbracket,\,\mathrm{and}\,
A \coloneqq C\, \backslash \, I.
\end{align*}
The nomenclature is inspired from the traditional assignment of $i$ and $a$ indices for ``occupied'' ($\varphi _i$ with $i \in I$) and ``virtual'' ($\varphi _a$ with $a\in A$) spinorbitals respectively. The $C$ integer interval is then seen as the set of indices covering the total canonical space, hence allowing the definition of the so-called \textit{canonical} orthonormal family — that we will call the canonical ordered basis:
\begin{align}
\mathcal{B} &\coloneqq (\varphi _r)_{r\in C}\nonumber \\ &= (\chi_r \otimes \sigma_r)_{r\in C} \label{eq:CanonicalBasis}
\end{align}
and the corresponding basis set $\{ \varphi_r  :  (r\in C)\}$. Through all the paper we will also use the following nomenclature: $\textbf{0}_o$ will denote the $N\times N$ zero matrix, $\textbf{0}_{ov}$ the $N\times (L-N)$ zero matrix, $\textbf{0}_{vo}$ the $(L-N)\times N$ zero matrix, and $\textbf{0}_v$ the $(L- N) \times (L-N)$ zero matrix.

We will also assume that from any of the excited-state quantum chemical calculation method considered in this contribution we have produced a finite set of $M$ molecular electronic excited states. We then set
\begin{equation*}
S \coloneqq \llbracket 1, M\rrbracket .
\end{equation*}
For any electronic transition from $ \psi _0$, namely the ground electronic state — often named \textit{Fermi vacuum} in the literature — to an excited electronic state $\psi _m$ (with $m\in S$), when ``$0m$'' is used as a subscript or a superscript, it is intended to be representative of the $(\psi _0 \longrightarrow \psi _m)$ transition — see the related comment in ref. \cite{etienne_towards_2021} where it is explained that such an implicit convention is not universal at all. When there is no such ``$0m$'' subscript or superscript, the transition is implicitely supposed to be taking place from the ground electronic state to a higher-energy electronic steady state.

In this paper we abusively extend the use of the universal quantifier to definitions. We also use the logical connective of conjunction ``$ \land$'' in several propositions. This symbol is always contextually clearly distinguished from the outer product when both are present in the same expression. Dirac's \textit{bra}-\textit{ket} notation will be used as follows: let $f$ and $g$ be two functions on a real vector space $V$, and $\hat{A}$ be a linear transformation from $V$ to $V$. Then, we will understand the $\braket{f|\hat{A}|g}_V$ expression in Dirac's notation as $\braket{f, \hat{A}g}_V$, where $\braket{\cdot \,, \cdot}_V$ is the real inner product in $V$. Most of the time we omit to place the vector space symbol as a subscript to the inner product for the sake of simplicity. Through all the text, the ``$\top$'' superscript will denote the \textit{transpose} of a matrix, and should not be confused with the ``$\mathrm{T}$'' superscript, standing for ``transition'' — often used to denote the one-particle reduced \textit{transition} density matrix. Given a finite-dimensional symmetric matrix \textbf{A}, its spectrum — i.e., the set of its eigenvalues — will be denoted $\sigma(\textbf{A})$.

\section{Preliminary considerations}

\subsection{Basic results from linear algebra}

\noindent This subsection relates few elementary propositions about matrices that will be used in this paper. Since these results are very basic, and since they are not the object of this paper but will be used as instruments for our own derivations, they are just stated here without proof. Notice that these results are given in this section for complex matrices for the sake of generality. In this paper we will deal with matrices with real-valued entries. Hence the results presented here directly apply to any matrix of interest to us in the present contribution, whose transpose can be regarded as being readily its adjoint.

From the existence of a singular value decomposition for any matrix, we arrive at some propositions that will be useful for deriving properties of matrices of interest for us in this study. The first definition relates the singular value decomposition (SVD) of a matrix:
\begin{definition}\label{theo:SVDvalues}
Let ${\normalfont \textbf{A}}$ be an $m\times n$ complex matrix and ${\normalfont \textbf{A}}^\dag$ its adjoint. Let $p$ be any strictly positive integer lower or equal to $\mathrm{min}\left(m,n\right)$. A non-negative real number $s_p$ is a singular value of ${\normalfont \textbf{A}}$ if and only if there exists a unit vector $\normalfont{\textbf{w}_p} \in \mathbb{C}^{m\times 1}$ — termed left-singular vector of $\normalfont{\textbf{A}}$ —, and a unit vector $\normalfont{\textbf{v}_p} \in \mathbb{C}^{n\times 1}$ — termed right-singular vector of $\normalfont{\textbf{A}}$ —, such that
\begin{align*}
  (\normalfont{\textbf{A}\textbf{v}_p = s_p \textbf{w}_p})\; \land \;(\normalfont{\textbf{A}\!^\dag \textbf{w}_p = s _p \textbf{v}_p}).
\end{align*}
\end{definition}
\noindent From Definition \ref{theo:SVDvalues} we directly find 
\begin{lemma}\label{theo:eigenvalues}
Let ${\normalfont \textbf{A}}$ be an $m\times n$ complex matrix and ${\normalfont \textbf{A}}^\dag$ its adjoint. Let $q$ be the strictly positive integer equal to $\mathrm{min}\left(m,n\right)$. Then, ${\normalfont \textbf{AA}^\dag}$ and ${\normalfont\textbf{A}^\dag\textbf{A}}$ share $q$ non-negative eigenvalues. Those eigenvalues are the squared singular values of {$\normalfont \textbf{A}$}.
\end{lemma}
\noindent Notice that replacing ``$\min(m,n)$'' by ``$\mathrm{rank}(\textbf{A})$'' in Lemma \ref{theo:eigenvalues} would go together with replacing ``nonnegative eigenvalues'' by ``strictly positive eigenvalues''. From Lemma \ref{theo:eigenvalues} we directly find
\begin{corollary}\label{lemma:prodPD}
\textit{Let} $\displaystyle
{\normalfont \textbf{A}}$ \textit{be an $m\times n$ complex matrix and {\normalfont $\textbf{A}^\dag$} its adjoint. Then,}
\begin{equation*}
({\normalfont\textbf{A}\textbf{A}^\dag} \succeq 0) \; \land \;
({\normalfont\textbf{A}^\dag\textbf{A}} \succeq 0).
\end{equation*}
{\it In other words, the product of a matrix by its adjoint is a positive semi-definite matrix.}
\end{corollary}
\noindent From the definition of positive semi-definiteness for a matrix, we get the following lemma:
\begin{lemma}\label{lemma:sumPD}
\textit{Let} $\displaystyle
{\normalfont \textbf{A}}$ \textit{and} {\normalfont $\textbf{B}$} \textit{be two $n\times n$ complex matrices. Then,}
\begin{equation*}
\normalfont \left(\textbf{A} \succeq 0 \; \land \; \textbf{B} \succeq 0\right) \;\Longrightarrow \; \left(\textbf{A}+\textbf{B} \succeq 0\right),
\end{equation*}
{\it i.e., the sum of two positive semi-definite matrices is a positive semi-definite matrix.}
\end{lemma}

\subsection{Auxiliary functions}

Given that we will have to deal with auxiliary wave functions in this paper, and due to the fact that the ``auxiliary function'' expression is rather polysemic, we would like to provide explanations about what we mean when we use it in the context of molecular electronic transition analysis. Here is a tentative definition — in what follows, $\mathcal{I}$, $\mathcal{K}$, and $\mathcal{L}$ are three at most countable, ordered sets of integers.

\begin{definition}[System-$\mathbb{I}$]
Let $\mathbb{I} \coloneqq (\psi_i)_{i\in \mathcal{I}}$ be a list of functions. Let $\mathcal{F}^\mathbb{I} \coloneqq (f_k^\mathbb{I})_{k\in\mathcal{K}}^{\textcolor{white}{\mathbb{I}}}$ be a list of functions whose definition implies at least one element of $\mathbb{I}$. Let $\mathscr{F}^{\mathbb{I}}\coloneqq (F_l^\mathbb{I})_{l\in\mathcal{L}}^{\textcolor{white}{\mathbb{I}}}$ be a list of functionals whose definition implies at least one element of $\mathbb{I}$.
\end{definition}

\begin{definition}[System-$\mathbb{J}$]
Let $\mathbb{J} \coloneqq (\Psi_i)_{i\in \mathcal{I}}$ be a list of functions. Let $\mathcal{F}^{ \mathbb{J}} \coloneqq (f_k^{\mathbb{J}})_{k\in\mathcal{K}}^{\textcolor{white}{\mathbb{I}}}$ be the list of functions of $\mathcal{F}^\mathbb{I}$ in which, for every $i$ in $\mathcal{I}$, every occurrence of $\psi_i$ has been replaced by $\Psi_i$. Let $\mathscr{F}^{\mathbb{J}}\coloneqq (F_l^{ \mathbb{J}})_{l\in\mathcal{L}}^{\textcolor{white}{\mathbb{I}}}$ be the list of functionals of $\mathscr{F}^\mathbb{I}$ in which, for every $i$ in $\mathcal{I}$, every occurrence of $\psi_i$ has been replaced by $\Psi_i$.
\end{definition}

\begin{definition}[$\mathbb{I}$--auxiliary function]
A function $\Psi$ from $\mathbb{J}$ is said to be an $\mathbb{I}$--auxiliary function if, substituting every occurrence of a properly chosen element of $\mathbb{I}$ by $\Psi$ in the elements of both $\mathcal{F}^\mathbb{I}$ and $\mathscr{F}^\mathbb{I}$, one expects to obtain a list of functions and a list of functionals as close as possible from $\mathcal{F}^\mathbb{I}$ and $\mathscr{F}^\mathbb{I}$, respectively.
\end{definition}

\begin{example}\label{ex:2e}
Consider the electronic Hamiltonian of a two-electron system. Let $\psi _0$ and $\psi_1$ be two eigenstates of that Hamiltonian. The one-body reduced transition density function corresponding to the $(\psi_0 \longrightarrow \psi _1)$ transition is defined as
$$\normalfont{n^\mathrm{T}_{\psi_0\rightarrow \psi_1} \coloneqq [\textbf{r}_1 \longmapsto 2 \int _{\mathbb{R}^3} \mathrm{d}\textbf{r}_2\sum_{s_2\in \llbracket 1, 2\rrbracket} \sum _{s_1\in \llbracket 1, 2\rrbracket} \, \psi_1(\textbf{r}_1,s_1,\textbf{r}_2,{s}_2)\psi_0^*(\textbf{r}_1,s_1,\textbf{r}_2,{s}_2)]}.$$
Suppose we are confident in a method which produces a function that we know is the exact $\psi_0$ but we do not have a method for deriving the explicit expression of $\psi_1$. Suppose now that we have a method for deriving a function $\Psi_1$ such that we expect $\normalfont{n^\mathrm{T}_{\psi_0\rightarrow \Psi_1}}$ to be very close to the actual $\normalfont{n^\mathrm{T}_{\psi_0\rightarrow \psi_1}}$. The $\Psi_1$ function will then be said ``auxiliary'' in that context. 
\end{example}

\begin{example}\label{ex:2ebis}
Consider the $\psi_1$ two-electron state from example {\normalfont{\ref{ex:2e}}}. One example of functional of $\psi_1$ is the number of electrons of the system described by $\psi_1$, that we can actually write
$$\normalfont{N[\psi_1]} = \int _{\mathbb{R}^3} \mathrm{d}\textbf{r}_1 \, n^\mathrm{T}_{\psi_1\rightarrow \psi_1}(\textbf{r}_1)$$
and that is equal to two. When designing an auxiliary wave function $\Psi_1$ as in example {\normalfont{\ref{ex:2e}}} we can, for example, decide to impose the constraint that $N[\Psi_1] = N[\psi_1]$ — this is usually the case.
\end{example}
\noindent Comparing examples \ref{ex:2e} and \ref{ex:2ebis} we see that the design of auxiliary wave functions may be performed under constraints: auxiliary wave functions will be declared admissible because they meet certain criteria, like for example that the corresponding state one body reduced density function integrates to the number of electrons.

When dealing with a complex molecular system, the EOM--TDDFRT method does not produce excited eigenstates of the electronic Hamiltonian. Whilst certain properties — like the transition dipole moments — can readily be evaluated using objects that are part of the outcome of the EOM-TDDFRT calculations, some other properties — like non-adiabatic couplings — have been computed using some auxiliary wave functions specifically built for that purpose \cite{crespo-otero_recent_2018,plasser_surface_2014,tapavicza_trajectory_2007,tavernelli_non-adiabatic_2009,tavernelli_nonadiabatic_2009,tavernelli_nonadiabatic_2009-1,tavernelli_nonadiabatic_2010,wang_nactddft_2021,hu_nonadiabatic_2007,hu2010nonadiabatic}.

\section{Natural-orbital representation of molecular electronic transitions}

We first provide a very brief reminder of elementary results about the eigenvalue (respectively, singular value) decomposition of the one-body reduced difference (respectively, transition) density matrix. Many more details about the theoretical construction and practical evaluation of these matrices and of their decomposition are to be found in Refs. \cite{etienne_comprehensive_2021,etienne_towards_2021}.

\subsection{One-body reduced difference density matrices}

Under the hypotheses and conditions of section \ref{sec:hyp} we first recall the definition of the general $(\psi_0 \longrightarrow \psi_m)$ transition 1--DDM elements in \cite{etienne_comprehensive_2021,etienne_towards_2021} for an orthonormal basis of real-valued spinorbitals: 
$$\forall m\in S, \,\forall (r,s)\in C^2,\, \left({\bm{\gamma}}^\Delta_{0\rightarrow m}\right)_{s,r}= \braket{\psi _m | \hat{r}^\dag\hat{s}|\psi_m} - \braket{\psi _0 | \hat{r}^\dag\hat{s}|\psi_0},$$
where $\hat{r}^\dag$ is the second quantization creation operator corresponding to $\varphi_r$, and $\hat{s}$ is the second quantization annihilation operator corresponding to $\varphi_s$. 

Every ${\bm{\gamma}}^\Delta_{0\rightarrow m}$ is symmetric. Let ${\bm{\gamma}}^\Delta$ be any of them. It is the matrix representation in $\mathcal{B}$ of an operator — see more details in \cite{etienne_towards_2021}. Let 
\begin{equation*}
\sigma\!\left({\bm{\gamma}}^\Delta\right) = \left\lbrace e_p \, : \, p \in \llbracket 1,L\rrbracket\right\rbrace
\end{equation*}
be the set of its eigenvalues, and $\textbf{U} = (\textbf{u}_p)_{p\in\llbracket 1,L\rrbracket}$ be the $L-$tuple of its eigenvectors with, for every $p$ in $\llbracket 1,L\rrbracket$, the $p^\mathrm{th}$ eigenvector $\textbf{u}_p$ corresponding to eigenvalue $e_p$. 
\begin{definition}
The eigenvectors of the one-body reduced difference density matrix are named {\normalfont{natural difference orbitals}} {\normalfont{({NDO}'}}s{\normalfont{)}}, and the eigenvalues are named {\normalfont{transition occupation numbers}}. 
\end{definition}
\noindent The transition occupation numbers are sometimes called \textit{difference occupation numbers} due to their origin. We then define the detachment density matrix related to ${\bm{\gamma}}^\Delta$ as
\begin{equation}
\bm{\gamma}^d \coloneqq \textbf{U}\mathrm{diag}(-\min(e_p,0))_{p\in\llbracket 1,L\rrbracket}\textbf{U}\tmat
\end{equation}
and the attachment density matrix related to ${\bm{\gamma}}^\Delta$ as
\begin{equation}
\bm{\gamma}^a \coloneqq \textbf{U}\mathrm{diag}(\max(e_p,0))_{p\in\llbracket 1,L\rrbracket}\textbf{U}\tmat.
\end{equation}
Let $n^\Delta$ be the $\mathbb{R}^3 \longrightarrow \mathbb{R}$ map corresponding to $\bm{\gamma}^\Delta$ that is named \textit{one-body reduced difference density}: 
\begin{equation}\label{eq:1DDf}
n^\Delta \coloneqq [\textbf{r} \longmapsto n^\Delta(\textbf{r}) = \sum_{p=1}^L\sum_{q=1}^L \sum_{s\in\llbracket 1,2\rrbracket} (\bm{\gamma}^\Delta)_{p,q}\,\varphi_p(\textbf{r},s)\varphi_q(\textbf{r},s)].
\end{equation}
\begin{proposition}We simultaneously have
\begin{equation}
\normalfont{\sum_{i_1=1}^L \mathrm{min}(e_{i_1},0) + \sum_{i_2=1}^L \mathrm{max}(e_{i_2},0) = \int _{\mathbb{R}^3} \mathrm{d}\textbf{r}_1\, \mathrm{min}(n^\Delta(\textbf{r}_1),0) + \int _{\mathbb{R}^3} \mathrm{d}\textbf{r}_2\, \mathrm{max}(n^\Delta(\textbf{r}_2),0)}
\end{equation} 
\textit{which is equal to zero, but}
\begin{equation}
\normalfont{\sum_{i_1=1}^L \mathrm{min}(e_{i_1},0) \leq \int _{\mathbb{R}^3} \mathrm{d}\textbf{r}_1\, \mathrm{min}(n^\Delta(\textbf{r}_1),0),}
\end{equation}
\textit{and}
\begin{equation}
\normalfont{ \sum_{i_2=1}^L \mathrm{max}(e_{i_2},0) \geq \int _{\mathbb{R}^3} \mathrm{d}\textbf{r}_2\, \mathrm{max}(n^\Delta(\textbf{r}_2),0).}
\end{equation}
\end{proposition}
\noindent Proof is given in \cite{etienne_boundary}.

\subsection{One-body reduced transition density matrices}

On the other hand, the definition of the general $(\psi_0 \longrightarrow \psi_m)$ transition 1--TDM elements in \cite{etienne_comprehensive_2021,etienne_towards_2021} for our orthonormal ordered basis of real-valued spinorbitals $\mathcal{B}$, is 
$$\forall m\in S, \,\forall (r,s)\in C^2,\, \left({\bm{\gamma}}^\mathrm{T}_{0\rightarrow m}\right)_{s,r}= \braket{\psi _0 | \hat{r}^\dag\hat{s}|\psi_m}.$$
As we have done previously, we consider the 1--TDM corresponding to \textit{any} transition. This time it is not necessarily symmetric. Consequently we perform its SVD — see Definition \ref{theo:SVDvalues}. 
\begin{definition}
The left- and right-singular vectors of the one-body reduced transition density matrix are called the {\normalfont left-} and {\normalfont right-natural transition orbitals} {\normalfont{({NTO}'}}s{\normalfont{)}}.
\end{definition}

\section{A quick review of CIS(--like) many-body wavefunctions features}\label{sec:CIS}
In this section we will start by recalling some results from linear algebra, and by introducing the CIS(--like) wavefunctions, i.e., wavefunctions being defined as or having the same properties as CIS wavefunctions. We will then review some basic properties related to those particular wavefunctions.

\subsection{CIS(--like) wavefunctions}

\noindent We now examine the case of any single-reference electronic excited state wavefunction written as a linear combination of singly-excited determinant states. In $\mathcal{B}$, it reads
\begin{equation} \label{eq:CISwf}
\tilde{\psi}_m  ^\zeta  = \sum _{i=1}^N \sum _{a=N+1}^{L} (\textbf{c}_m)_{ia} \, \textcolor{black}{\hat{a}^\dag} \textcolor{black}{\hat{i}}\psi _0
\end{equation}
with $(m\in S)$ and $\textbf{c}_m \in \mathbb{R}^{N(L-N)\times 1}$. The $\textbf{c}_m$ vector in \eqref{eq:CISwf} is released during the excited-state calculation. In \eqref{eq:CISwf}, $\hat{i}$ is the second quantization operator annihilating an electron in spinorbital $\varphi _i$ and $\hat{a}^\dag$ is the second quantization operator creating an electron in spinorbital $\varphi _a$. If $\tilde{\psi}^\zeta _m$ is not normalized, we can write the corresponding normalized ansatz as $\psi ^\zeta _m = z^{-1/2}_{\zeta,m}\tilde{\psi}^\zeta_m$, with $$z_{\zeta,m} \, {\vcentcolon=} \,\textbf{c}_m^{\!\top}\!\textbf{c}_m^{\textcolor{white}{\dag}}.$$ 
\begin{definition}
Any ansatz that can be written as a linear combination of $N$-electron Fock states singly-excited relatively to the Fermi vacuum will be called a $\zeta$--ansatz (or $\zeta$--type ansatz). 
\end{definition}

\subsection{Properties of $\zeta$--type ansatzes}\label{par:PropZetaansatzes}

\begin{definition} We denote by $W^\zeta$ the set of all $\zeta$--type ansatzes.\end{definition}
\noindent Any element of $W^\zeta$ being an $N$--electron state described using an $L$--dimensional orthonormal basis can be related, together with its reference state, with the corresponding 1--TDM and 1--DDM, as well as other objects. All those will have the properties depicted in the present subsection.

\subsubsection{$\zeta$-type one-body reduced transition density matrix}

\paragraph{{\thesubsubsection}.A. Structure}$\;$ \\

\noindent Since every $\hat{a}^\dag\hat{i}\psi_0$ configuration in \eqref{eq:CISwf} is orthogonal to $\psi _0$ we find, using the definition of the general ($\psi_0 \longrightarrow \psi_m$) 1--TDM elements for an orthonormal basis of real-valued spinorbitals — \textit{vide supra} — that, for our trial wavefunction $\tilde{\psi}^\zeta_m$ we have that, in $\mathcal{B}$,
\begin{align}\label{eq:1TDMstrX}
\forall (i,a) \in I\times A, \quad &(\tilde{{{\bm{\gamma}}}}^\mathrm{T}_{\zeta,0\rightarrow m} )_{a,i} = (\textbf{c}_m)_{ia},
\\\label{eq:1TDMstr0a}
\forall (s,r)\in C^2\, \backslash \, (I\times A), \quad &(\tilde{{{\bm{\gamma}}}}^\mathrm{T}_{\zeta,0\rightarrow m} )_{s,r} = 0.  
\end{align}
For the sake of readability, we will consider ``any'' electronic transition — i.e., any among the $M$ available in our description — so we will drop the ``$0\rightarrow m$'' subscripts through all this subsection.

We define the 1--TDM corresponding to such a $\zeta$--ansatz as ${{\bm{\gamma}}}^\mathrm{T}_{\zeta }  \coloneqq z_\zeta^{-1/2} {\tilde{\bm{\gamma}}}^\mathrm{T}_\zeta$. We see from \eqref{eq:1TDMstrX} and \eqref{eq:1TDMstr0a} that ${{\bm{\gamma}}}^{\mathrm{T}}_{\zeta }$ is partitioned as
\begin{eqnarray}\label{eq:gammaTzeta}
{{\bm{\gamma}}}^\mathrm{T}_\zeta = \left( 
 \begin{array}{cc}
\textbf{0}_{o\textcolor{white}{v}}  & \textbf{0}_{\textcolor{black}{ov}}   \\
\textbf{T}\tmat_{\textcolor{white}{v}}  &  \textbf{0}_v \\
 \end{array}\right) 
\end{eqnarray} 
with $\textbf{T}$ an $(N\times(L-N))$ real matrix and, for every $(i,a)$ in $I\times A$, $z_\zeta^{-1/2}\textbf{c}_{ia} = (\bm{\gamma}^\mathrm{T}_\zeta)_{a,i}$ and $ (\bm{\gamma}^\mathrm{T}_\zeta)_{a,i}= (\textbf{T})_{i,a-N}$. The question of the normalization is absent in the CIS formalism because the ansatz is usually readily normalized there, but we will see below that some ansatzes that we will manipulate are not normalized in their native form.

\paragraph{{\thesubsubsection}.B. Singular value decomposition}$\;$ \\

\noindent We are now going to review few properties mostly related with the singular value decomposition of the $\zeta-$type one-body reduced transition density matrix.
\begin{proposition}
Any $\zeta$--type {\normalfont 1--TDM} is nilpotent at every order.
\end{proposition}
\begin{proof}
\noindent Multiplying the ${{\bm{\gamma}}}^\mathrm{T}_{\zeta }$ above by itself yields the $L\times L$ zero matrix $\textbf{0}_L$. Given that the second power of any $\zeta$--type 1--TDM is equal to $\textbf{0}_L$, any $n^\mathrm{th}$ power (with $n$ a positive integer superior to two) of ${\bm{\gamma}}^\mathrm{T}_{\zeta }$ will also be equal to $\textbf{0}_L$.
\end{proof}
\noindent If we write ${{\bm{\gamma}}}^\mathrm{T}_{\zeta }$ as the $L$--tuple of its columns, i.e., $\bm{\gamma}^\mathrm{T}_{\zeta } = (\textbf{g}_r)_{r\in C}$, we find that, for every $r$ in $C$, $\textbf{g}_r\in \mathrm{ker}\left({{\bm{\gamma}}}^\mathrm{T}_\zeta \right)$, where the kernel of ${{\bm{\gamma}}}^\mathrm{T}_\zeta$ is defined as $$\mathrm{ker}\left({{\bm{\gamma}}}^\mathrm{T}_{\zeta}\right) \coloneqq \left\lbrace \textbf{w} \in \mathbb{R}^{L\times 1} \, : \, \bm{\gamma}^\mathrm{T}_{\zeta} \textbf{w} = \textbf{0}_L\right\rbrace.$$ This means that the columns of $\bm{\gamma}^{\mathrm{T}}_{\zeta}$ are column vectors that all belong to the kernel of $\bm{\gamma}^{\mathrm{T}}_{\zeta}$. Due to the nilpotency of $\zeta$--type 1--TDM's, one rather performs a singular value decomposition (SVD) of its southwest block, the rectangular $\textbf{T}^\top$. The SVD of $\textbf{T}^\top$ reads
\begin{align} 
\exists (\textcolor{black}{\textbf{O}}, \textbf{V}, \bm{\lambda}) \in \mathbb{R}^{N\times N} \times \mathbb{R}^{(L-N)\times(L-N)} \times \mathbb{R}_+^{N\times (L-N)},    \textbf{T}^\top  = \textcolor{black}{\textbf{V}}\bm{\lambda}\tmat   \textcolor{black}{\textbf{O}}\tmat \label{eq:svdntos}
\end{align}
with $\displaystyle \textbf{O}\tmat = \textbf{O}^{-1}$ and $\displaystyle \textbf{V}\tmat = \textbf{V}^{-1}$. The $\bm{\lambda}$ matrix satisfies 
$$\forall (p,p') \in  I \times \llbracket 1,(L-N)\rrbracket, \, (p \neq p') \, \Longrightarrow \, (\bm{\lambda})_{p,p'} = 0.$$
We then set $q \coloneqq \mathrm{min}\left(N,(L-N)\right)$ and, for every $p$ in $\llbracket 1,  q\rrbracket$, we define $$ \lambda _p \coloneqq (\bm{\lambda})_{p,p}.$$ With this in hand, the SVD of ${{\bm{\gamma}}}^\mathrm{T}_\zeta$ is then found to be ${{\bm{\gamma}}}^\mathrm{T}_\zeta = \textbf{Q}\bm{\Sigma}\textbf{P}\tmat,$ with
\begin{equation}\label{eq:QSigmaP}
\textbf{Q} =  \left(\begin{array}{cc}
\textbf{0}_{ov}  & \textbf{0}_{o{\textcolor{white}{o}}}   \\
\textbf{V}  &  \textbf{0}_{vo} \\
 \end{array}\right) \!, \; \textbf{P} = \textbf{O} \oplus \textbf{0}_{v},\;\mathrm{and}\; \bm{\Sigma} = \bm{\lambda}\tmat\oplus\textbf{0}_{ov}.
\end{equation}
Columns of \textbf{V} contain the components of the so-called left ($v$) NTOs in the $(\varphi _r)_{r\in A}$ ordered basis, and the columns of \textbf{O} contain the components of the right ($o$) NTOs in the $(\varphi _r)_{r\in I}$ ordered basis: for every $p$ in $\llbracket 1,q \rrbracket $, we build
\begin{align*}
\varphi ^o_p  = \sum _{l=1}^N (\textbf{O})_{l,p} \,\varphi _l \quad \mathrm{and} \quad
\varphi ^v_p  = \sum _{l=1}^{L-N} (\textbf{V})_{l,p}\, \varphi _{N+l}^{\textcolor{white}{\dag}}  .
\end{align*}
Accordingly, we see that there exists an exact and unique pairing between left and right NTOs through their singular values, so that the canonical picture of a linear combination of singly-excited configurations in the $\mathcal{B}$ ordered basis can be transposed in the NTOs basis without approximation. Given that the problem is originally expressed in a finite-dimensional local basis of $L$ one-body wavefunctions, the NTO picture is the most compact representation of the electronic transition — the expansion is then reduced from $N\times (L-N)$ terms to $\min(N,(L-N))$ terms — with the exact conservation of the $\zeta$--ansatz nature. According to \eqref{eq:svdntos}, we also have that
\begin{align}\label{eq:OTTO}
\textbf{O}\tmat \textbf{TT}\tmat \textbf{O} &= \bm{\lambda}\bm{\lambda}\tmat \in \mathbb{R}_+^{N\times N},
\\ \label{eq:VTTV}
\textbf{V}\tmat \textbf{T}\tmat\textbf{T} \textbf{V} &= \bm{\lambda}\tmat\bm{\lambda} \in \mathbb{R}_+^{(L-N)\times (L-N)}.
\end{align}

\subsubsection{$\zeta$-type one-body reduced difference density matrix}

\paragraph{{\thesubsubsection}.A. Structure} $\;$ \\

\noindent Combining Corollary \ref{lemma:prodPD} with time-independent Wick's theorem for fermions we find
\begin{proposition} 
$\zeta$--type {\normalfont 1--DDM} is the direct sum of a negative semidefinite and a positive semidefinite matrix.
\end{proposition}
\noindent This was originally stated in ref. \cite{head-gordon_analysis_1995}. Complete proof is given in ref. \cite{etienne_comprehensive_2021} where the following structure for $\textcolor{black}{\bm{\gamma} ^\Delta_\zeta}$ was derived: In $\mathcal{B}$, we have
\begin{equation}\label{eq:structgammaDelta}
\textcolor{black}{\bm{\gamma} ^\Delta_\zeta} = \left(- \textbf{TT}\tmat\mkern3mu\right) \oplus \textbf{T}\tmat\textbf{T}  . 
\end{equation}

\paragraph{{\thesubsubsection}.B. Spectral properties — i. Generalities, detachment/attachment} $\;$

\noindent Using again Corollary \ref{lemma:prodPD} we find
\begin{corollary} \label{cor:zetaWithoutDiag}
$\zeta$--type detachment and attachment density matrices can be derived from the transition density matrix without matrix diagonalisation.
\end{corollary} 
\begin{proof}
\noindent Due to the definition of the detachment and attachment density matrices in terms of the sign of the transition occupation numbers, i.e., eigenvalues of the one-body reduced difference density matrix, we find that, according to Corollary \ref{lemma:prodPD},
\begin{align*}
\textbf{TT}\tmat \oplus \textbf{0}_{\textcolor{black}{v}}&= \textcolor{black}{\bm{\gamma} ^{d}_\zeta},\\
  \textbf{0}_{\textcolor{black}{o}} \oplus \textbf{T}\tmat\textbf{T} &= \textcolor{black}{\bm{\gamma} ^a_\zeta}. \tag*{\qedhere}
\end{align*}
\end{proof}
\noindent Notice that Corollary \ref{cor:zetaWithoutDiag} was also originally stated in other words in ref. \cite{head-gordon_analysis_1995} where the detachment/attachment representation of molecular electronic transitions was first introduced in the context of CIS calculations. 
\begin{proposition}\label{prop:CIS-NTOsNDOs} In $\mathcal{B}$, the $\zeta$-natural difference and transition orbitals are identical.
\end{proposition}
\begin{proof}
\noindent Due to the structure of $\textcolor{black}{\bm{\gamma} ^\Delta _\zeta}$ in \eqref{eq:structgammaDelta}, we have the following possible matrix of eigenvectors
\begin{equation}\label{eq:MOV}
\textbf{U} = {\textcolor{black}{\textbf{O}}} \oplus {\textcolor{black}{\textbf{V}}} 
\end{equation}
and, given \eqref{eq:OTTO} and \eqref{eq:VTTV}, we find the following possible eigenvalue decomposition: 
\begin{equation*}
\textbf{U}\tmat  \textcolor{black}{\bm{\gamma} ^\Delta _\zeta} \textbf{U}  = \left(- \bm{\lambda}\bm{\lambda}\tmat\, \right) \oplus \bm{\lambda}\tmat\bm{\lambda}
\end{equation*}
which means that the CIS natural difference orbitals are paired just as the natural transition orbitals are. In fact, according to \eqref{eq:QSigmaP} and \eqref{eq:MOV}, their components in $\mathcal{B}$ are identical.
\end{proof}
\noindent We finally establish that the trace of $\zeta$--type detachment and attachment 1--RDM's, i.e., $\vartheta _\zeta$, is equal to
\begin{equation*}
 \vartheta _\zeta = \mathrm{tr}\left(\bm{\lambda \lambda}\tmat\, \right) = \mathrm{tr}\left(\bm{\lambda\tmat \lambda} \right) = \sum _{p=1}^q \lambda _p^2.
 \end{equation*} 
 
\paragraph{{\thesubsubsection}.B. Spectral properties — ii. Eigenvalues numbering} $\;$ \\
 
\begin{definition}[Inertia of a complex matrix \cite{de1981inertia}] Let n be a non-zero natural integer. \textit{Let} {\normalfont{\textbf{M}}} \textit{be a complex, square matrix of order $n$. The {\normalfont{inertia}} of} {\normalfont{\textbf{M}}}, noted {\normalfont{$\mathrm{In}(\textbf{M})$}}, is the ordered triple of natural integers
\begin{equation*}
\mathrm{In}({\normalfont{\textbf{M}}}) = (\pi_M,\nu_M,\delta_M),
\end{equation*}
\textit{where} $\pi_M$ \textit{is the number of eigenvalues of} {\normalfont{\textbf{M}}} \textit{with a strictly positive real part}, $\nu_M$ \textit{is the number of eigenvalues of} {\normalfont{\textbf{M}}} \textit{with a strictly negative real part, and} $\delta_M$ \textit{is the number of eigenvalues of} {\normalfont{\textbf{M}}} with a zero real part. In particular, if {\normalfont \textbf{M}} is Hermitian, $\pi_M$ {\normalfont(}respectively, $\nu_M)$ \textit{is the number of its strictly positive {\normalfont(}respectively, strictly negative{\normalfont)} eigenvalues.}
\end{definition}
\noindent In the following, the number of strictly positive — respectively, negative — eigenvalues of a given Hermitian matrix $\textbf{M}$ will sometimes be denoted $\pi(\textbf{M})$ — respectively, $\nu(\textbf{M})$.

We would like to derive boundary values for the elements of $\mathrm{In}(\bm{\gamma}^\Delta_\zeta)$ — i.e., boundary values for the number of strictly positive and strictly negative eigenvalues of $\bm{\gamma}^\Delta_\zeta$. The result — rather expectable — will have to be compared with the one for the (un)relaxed EOM-TDDFRT one-body reduced difference density matrix. 

\begin{proposition}\label{prop:boundaryzeta}
Let $\bm{\gamma}^\Delta_\zeta$ be the $L\times L$ one-body reduced difference density matrix, for an $N$--electron system, corresponding to the $(\psi _0 \longrightarrow \psi)$ transition to a $\zeta$--state $(\psi \in W^\zeta)$ from its relative one-determinant reference state $\psi_0$. Let $(\pi_\zeta, \nu_\zeta, \delta_\zeta)$ denote the inertia of $\bm{\gamma}^\Delta_\zeta$. Then,
$$\pi_\zeta = \nu_\zeta \leq \min (N,(L-N)).$$
\end{proposition}
\begin{proof}
\noindent  From the structure of $\bm{\gamma}^\Delta_\zeta$, i.e., $(-\textbf{TT}\tmat) \oplus \textbf{T}\tmat\textbf{T}$, we readily see that $\pi_\zeta = \pi(\textbf{T}\tmat\textbf{T})$ and that $\nu_\zeta = \nu(-\textbf{TT}\tmat)$. From Lemma \ref{theo:eigenvalues}, we see that both numbers are lower or equal to $\min(N,(L-N))$. Using Lemma \ref{theo:eigenvalues} again, and due to the structure of $\bm{\gamma}^\Delta_\zeta$, we immediately see that
$$\forall e \in \sigma(\bm{\gamma}^\Delta_\zeta), \, (e \neq 0) \Longrightarrow \exists e ' \in \sigma (\bm{\gamma}^\Delta_\zeta), \, e ' = - e$$
and immediately deduce the desired result.
\end{proof} 
 
\paragraph{{\thesubsubsection}.B. Spectral properties — iii. Eigenvalues ordering} $\;$ \\
  
\noindent Before dealing with the ordering of the eigenvalues, we re-define the \textbf{O} and \textbf{V} matrices as two tuples of vectors, i.e., $\textbf{O} = (\textbf{o}_1, \dotsc , \textbf{o}_N)$ and $\textbf{V} = (\textbf{v}_1, \dotsc , \textbf{v}_{L-N})$. We also give the two following definitions:

\begin{definition} An $n$--tuple $(a_1^\downarrow, \dotsc , a_n^\downarrow)$ of real numbers is given in decreasing order if
\begin{equation*}
a_1^\downarrow \geq \cdots \geq a_n^\downarrow.
\end{equation*}
\end{definition}
\begin{definition}
An $n$--tuple $(a_1^\uparrow, \dotsc , a_n^\uparrow)$ of real numbers is given in increasing order if
\begin{equation*}
a_1^\uparrow \leq \cdots \leq a_n^\uparrow.
\end{equation*} 
\end{definition}
\noindent In our study of $N$--electron systems using an $L$--dimensional basis, three possibilities can be encountered: \textit{Case 1}, where ($N< (L-N)$); \textit{Case 2}, where ($N > (L-N)$); \textit{Case 3}, where ($N = (L-N)$). Since the theoretical and computational manipulations are usually performed within the framework characteristic of \textit{Case 1}, we put \textit{Case 2} and \textit{Case 3} in Appendix.

$\;$

\noindent \textit{Case 1}: $N< (L-N)$ — We define, for any ordering of the $(\lambda_p^2)_{p\in I}^{\textcolor{white}{a}}$ $N$--tuple,
\begin{equation*}
\bm{\Lambda} \coloneqq \bm{\lambda\lambda}\tmat\in \mathbb{R}_+^{N\times N},
\end{equation*}
i.e., $
\mathrm{diag}(\Lambda _1 , \dotsc , \Lambda _N) = \mathrm{diag} (\lambda _1 ^2 , \dotsc , \lambda _N^2)$. We have that $
\bm{\lambda}\tmat \bm{\lambda} = \bm{\Lambda} \oplus \textbf{0}_w$, with $\textbf{0}_w$ being the $(L-2N)\times(L-2N)$ zero matrix. According to
\begin{equation*}
\forall p \in \llbracket N+1,   L-N\rrbracket, \, \textbf{v}_p \in \mathrm{ker}\left(\textbf{T}\right),
\end{equation*}
where the kernel of \textbf{T} is defined as
\begin{equation*}
\mathrm{ker}\left(\textbf{T}\right) \coloneqq \left\lbrace \textbf{w} \in \mathbb{R}^{(L-N)\times 1} \, : \, \textbf{T}\textbf{w} = \textbf{0}_N\right\rbrace,
\end{equation*}
with $\textbf{0}_N$ being the $N\times 1$ zero column vector, we define two matrices: the first one is the $\textbf{V}^\uparrow _1 \coloneqq (\textbf{v}_1^\uparrow, \dotsc , \textbf{v}_N^\uparrow)$ $N$--tuple of vectors in which the column vectors are ordered such that our solution to the matrix eigenvalue problem for $\textbf{T}\tmat\textbf{T}$ can be rewritten in a way that makes the eigenvalues of $\textbf{T}\tmat\textbf{T}$ sorted in the increasing order, i.e., $\textbf{T}\tmat\textbf{T}\textbf{V}^\uparrow _1 = \textbf{V}^\uparrow_1\bm{\Lambda}^\uparrow$, where
\begin{equation*}
\bm{\Lambda}^\uparrow \coloneqq \mathrm{diag}\left(\Lambda _1 ^\uparrow , \dotsc , \Lambda _{N}^\uparrow\right).
\end{equation*}
The second matrix is simply $
\textbf{V}_{\mathrm{ker}} \coloneqq \left(\textbf{v}_{N+1}, \dotsc , \textbf{v}_{L-N}\right).$ While we identify $\textbf{V}$ as $\left( \textbf{V}_{\!O},\textbf{V}_{\mathrm{ker}}\right)$ where $\textbf{V}_{\!O}$ is the $(\textbf{v}_i)_{i\in \llbracket 1, N \rrbracket}$ $N$--tuple of column vectors, we can permute columns of \textbf{V} so that the column vectors belonging to $\mathrm{ker}(\textbf{T})$ appear first, i.e., $\textbf{V}_{\!>} \coloneqq  ( \textbf{V}_{\mathrm{ker}},\textbf{V}^\uparrow_1)$. We then have that our solution to the matrix eigenvalue problem for $\textbf{T}\tmat\textbf{T}$ can be rewritten in a way that makes the eigenvalues of $\textbf{T}\tmat\textbf{T}$ sorted in the increasing order:
\begin{equation*}
\left(\textbf{T}\tmat\textbf{T}\right)\textbf{V}_{\!>} = \textbf{V}_{\!>} \left(\textbf{0}_w \oplus \bm{\Lambda}^\uparrow\right).
\end{equation*}
We also define $\textbf{O}^\uparrow_1$ and $\textbf{O}^\downarrow_1$ such that, for every $\omega$ in $\left\lbrace \uparrow, \downarrow \right\rbrace$, we have $\textbf{TT}\tmat \textbf{O}_1^\omega = \textbf{O}^\omega_1 \bm{\Lambda}^\omega$. We finally define $\textbf{V}^\downarrow_1$ such that
\begin{equation*}
\textbf{T}\tmat\textbf{T}\textbf{V}^\downarrow_1 = \textbf{V}^\downarrow_1 \!\left(\bm{\Lambda}^\downarrow \oplus \textbf{0}_w\right).
\end{equation*}
This allows us to write the matrix of eigenvalues of $\bm{\gamma}_\zeta ^\Delta$ sorted in a decreasing order as
\begin{equation*}
\textbf{K}^\downarrow_1 = \bm{\Lambda}^\downarrow \oplus \textbf{0}_w \oplus \left(-\bm{\Lambda}^\uparrow\right)
\end{equation*}
with the corresponding matrix of eigenvectors:
\begin{eqnarray*}
\textbf{M}^\downarrow _1 \coloneqq  \left(
\begin{array}{cc}
\textbf{0}_{o\textcolor{white}{v}}  &\textbf{O}^\uparrow_1     \\
 \textbf{V}^\downarrow_1 &  \textbf{0}_v 
 \end{array}\right).
\end{eqnarray*}
On the other hand, the matrix of eigenvalues of $\bm{\gamma}_\zeta^\Delta$ sorted in a increasing order reads
\begin{equation*}
\textbf{K}^\uparrow_1 = \left(- \bm{\Lambda}^\downarrow \right) \oplus \textbf{0}_w \oplus \bm{\Lambda}^\uparrow
\end{equation*}
with the corresponding matrix of eigenvectors being $
\textbf{M}^\uparrow_1 = \textbf{O}^\downarrow_1 \oplus \textbf{V}_{\!>}.$

\section{Comparison of the EOM and auxiliary wavefunction pictures}

\noindent This section introduces and compares the two pictures (EOM \cite{ipatov_excited-state_2009,rowe_equations--motion_1968} and AMBW — with multiple variants) one meets when dealing with TDDFRT.

\subsection{EOM with an exact state-transfer operator} 

\noindent Let $\hat{{T}}_{0\rightarrow m}$ be a nilpotent, number- and norm-conserving transition operator corresponding to the transition between the $\psi_0$ and the $\psi _m$ quantum states (with $m \in S$):
\begin{equation*}
\hat{{T}}_{0\rightarrow m} = \ket{\psi _m}\bra{\psi _0}, \; \hat{{T}}^\dag_{0\rightarrow m} =  \ket{\psi _0}\bra{\psi _m} 
\end{equation*} 
with the constraint that the two states are normalized and orthogonal to each other. In these conditions, we can assess the equation-of-motion 1--DDM and 1--TDM elements: for every $(r,s)$ in $C^2$, the corresponding 1--DDM element expression reads
\begin{align*}
\left\langle \!\psi _0 \! \left| \left[ \hat{{T}}_{0\rightarrow m}^\dag  , \left[ \hat{r}^\dag \hat{s},\hat{{T}}_{0\rightarrow m}\right]\right] \right| \! \psi _0 \! \right\rangle &= \left\langle \!\psi _m \! \left| \hat{r}^\dag \hat{s}\right| \! \psi _m \! \right\rangle - \left\langle \!\psi _0 \! \left| \hat{r}^\dag \hat{s}\right| \! \psi _0 \! \right\rangle \nonumber \\ &= (\bm{\gamma}^\Delta_{0\rightarrow m})_{s,r}
\end{align*}
while the corresponding 1--TDM element expression reads 
\begin{equation}\label{eq:TopMatEl}
(\bm{\gamma}_{0\rightarrow m}^\mathrm{T} )_{s,r} = \left\langle \!\psi _0 \! \left|\left[ \hat{r}^\dag\hat{s},\hat{{T}}_{0\rightarrow m}\right]  \right| \! \psi _0 \! \right\rangle \! .
\end{equation} 
\subsection{The case of EOM-TDDFRT}
\noindent In the EOM formulation (denoted by a ``$\eta$'' symbol) of TDHF, the transition operator $\hat{{T}}$ in the derivation of the matrix elements — see \eqref{eq:TopMatEl} — is substituted by
\begin{equation*}
\hat{{T}}_\eta \coloneqq \sum _{i=1}^N \sum _{a=N+1}^L \left(\textbf{x}_{ia} \,\hat{a}^\dag\hat{i} - \textbf{y}_{ia} \,\hat{i}^\dag\hat{a}\right),  
\end{equation*} 
i.e., a truncated CI excitation operator corrected by deexcitation contributions. Note that $\hat{{T}}_\eta$ is not used as a direct approximation to an EOM transition operator $\ket{\psi _\eta}\!\bra{\psi _0}$ to write an excited state ansatz. Indeed, $\hat{{T}}_\eta$ is not nilpotent, not number-conserving, and not norm-conserving.

Here again, for the sake of readability of the equations, we will consider a general electronic transition between the ground state and any excited state of a molecule whose electron system is described by an electronic Hamiltonian $\hat{H}$. The components of the \textbf{x} and \textbf{y} vectors, solutions to the response equation
\begin{equation}
 \left(\textcolor{black}{ \begin{array}{cc}
\textbf{A}^{\textcolor{white}{*}}  & \textbf{B}  \\ \textbf{B}^{\textcolor{white}{*}}  & \textbf{A} 
\end{array}}\right) \left(\begin{array}{c}
\textcolor{black}{{\textbf{x}}} \\ \textcolor{black}{{\textbf{y}}}
\end{array}\right) = \textcolor{black}{\omega}  \left( \begin{array}{cc}
\textbf{I} & \textcolor{white}{-}\textbf{0} \\ \textbf{0} & -\textbf{I}
\end{array}\right) \left(\begin{array}{c}
\textcolor{black}{{\textbf{x}}} \\ \textcolor{black}{{\textbf{y}}}
\end{array}\right), \label{eq:casida}
\end{equation} 
with, for every $(i,j)$ in $I^2$ and every $(a,b)$ in $A^2$,
\begin{align}\label{eq:Amatrix}
\left(\textbf{A}\right)_{ia,jb} &= \left\langle \! \psi _0 \! \left| \left[ \hat{i}^\dag \hat{a},\left[\hat{H},\hat{b}^\dag\hat{j}\right]\right]\right| \! \psi _0 \! \right\rangle,\\
\label{eq:Bmatrix}
\left(\textbf{B}\right)_{ia,jb} &= \left\langle \! \psi _0 \! \left| \left[ \hat{i}^\dag \hat{a},\left[\hat{H},\hat{j}^\dag\hat{b}\right]\right]\right| \! \psi _0 \! \right\rangle,
\end{align}
can be recast into two real $(N\times (L-N))$ rectangular matrices:
\begin{align}
\forall (i,a) \in I \times A, \quad \textbf{x}_{ia} & = (\textbf{X})_{i,a-N},\\
  \textbf{y}_{ia} &= (\textbf{Y})_{i,a-N}.\label{eq:endrecastsquare}
\end{align}
Note that \textbf{x} and \textbf{y} are chosen to satisfy 
\begin{equation}
\textbf{x}\tvec\textbf{x}-\textbf{y}\tvec\textbf{y} = 1. \label{eq:x2-y2}
\end{equation}
In (\ref{eq:casida}), $\omega$ is the transition energy between the ground state and the excited state of interest and $\hat{H}$ is the electronic Hamiltonian. $\textbf{I}$ and $\textbf{0}$ are the $[N\times (L-N)]\times[N\times(L-N)]$ identity and zero matrices, respectively. In (\ref{eq:Amatrix}) and (\ref{eq:Bmatrix}), $ia$ and $jb$ are assimilated to individual integers (as it is done for indexing the components of \textbf{x} and \textbf{y}), for pointing the row (on the left of the comma) and column (on the right of the comma) indices of matrices \textbf{A} and \textbf{B} — see the introductory comments of this paper.

The EOM--TDDFRT central equation has the exact same structure as \eqref{eq:casida}. This comment also holds for the matrices introduced in the following subsection. This also stands for the Bethe-Salpeter excited-state claculation method \cite{blase_bethesalpeter_2018}. Hence, the structure and properties of the objects discussed below are identical in the three methods.
\subsubsection{Generalities about the EOM--TDDFRT matrices}
In the conditions detailed above, the transition density matrix reads
\begin{eqnarray}\label{eq:EOM-TDM}
  \left( 
 \begin{array}{cc}
\textbf{0}_o  & \textbf{Y}   \\
\textbf{X}\tmat\!  &  \textbf{0}_v \\
 \end{array}\right) = \bm{\gamma} ^{\mathrm{T}}_\eta \in \mathbb{R}^{L \times L}. 
\end{eqnarray}
Its transpose is often seen in the litterature to denote the same transition — here, ``$\mathrm{T}$'' denotes an electronic transition and ``$\top$'' denotes the transpose of a matrix, as stated at the beginning of this paper. However we have showed and discussed in refs. \cite{etienne_comprehensive_2021,etienne_towards_2021} that the transpose of the matrix given above corresponds to the transition between the two same states but with the departure/arrival states being permuted.
\begin{proposition}\label{prop:1DDM-directsum} The {\normalfont EOM--TDDFRT} one-body reduced difference density matrix is the direct sum of a negative semidefinite matrix and a positive semidefinite matrix.
\end{proposition}
\begin{proof}\noindent In the conditions detailed above, the difference density matrix is \cite{etienne_comprehensive_2021}
\begin{equation}\label{eq:EOM-DDM}
  \left(-\textbf{X}{\textbf{X}}\tmat - \textbf{Y}\textbf{Y}\tmat\, \right) \oplus \left({\textbf{X}}\tmat{\textbf{X}} + {\textbf{Y}}\tmat{\textbf{Y}} \right) =  \bm{\gamma}^\Delta_\eta \in \mathbb{R}^{L \times L}  
\end{equation} 
Hence, we see that the 1--TDM and 1--DDM in TDDFRT and in the Bethe-Salpeter method have identical structures to those in equations (\ref{eq:EOM-TDM}) and (\ref{eq:EOM-DDM}), respectively \cite{etienne_comprehensive_2021}. According to Lemma \ref{lemma:sumPD}, we see that
\begin{align*}
- \textcolor{black}{\left(\textbf{X}{\textbf{X}}\tmat + \textbf{Y}\textbf{Y}\tmat\, \right)} &\preceq 0, \\
\textcolor{white}{--}\textcolor{black}{\left({\textbf{X}}\tmat{\textbf{X}} + {\textbf{Y}}\tmat{\textbf{Y}} \right)} &\succeq 0,
\end{align*}
which concludes the proof.
\end{proof}
\begin{corollary}
The {\normalfont EOM--TDDFRT} detachment and attachment density matrices can be derived without matrix diagonalization. 
\end{corollary}
\begin{proof}
\noindent From the proof of Proposition \ref{prop:1DDM-directsum}, and according to the definition of the detachment and attachment, we can now give the expression of the EOM detachment and attachment one-body reduced density matrices:
\begin{align*}
\textcolor{black}{\left(\textbf{X}{\textbf{X}}\tmat + \textbf{Y}\textbf{Y}\tmat\, \right)} \oplus \textbf{0}_{\textcolor{black}{v}}&= \textcolor{black}{\bm{\gamma} ^{d} _\eta}, \\
 \textbf{0}_{\textcolor{black}{o}} \oplus \textcolor{black}{\left({\textbf{X}}\tmat{\textbf{X}} + {\textbf{Y}}\tmat{\textbf{Y}} \right)} &= \textcolor{black}{\bm{\gamma} ^a _\eta} .
\end{align*}
Deriving these two matrices can be done without requiring any matrix diagonalization.
\end{proof}

\subsubsection{$\eta$--type 1--DDM: eigenvalues numbering}

\begin{lemma}\label{lemma:H1plusH2}
Let $i$ be an integer belonging to $\llbracket 1,2\rrbracket$. Let $\pi,\nu,\delta,\pi_i,\nu_i$, and $\delta_i$ be nonnegative integers. The following statements are equivalent:

\noindent (a) There exist two $n\times n$ Hermitian matrices, ${\normalfont{\textbf{H}_1}}$ and ${\normalfont{\textbf{H}_2}}$ such that ${\normalfont \mathrm{In}(\textbf{H}_1+\textbf{H}_2) = (\pi, \nu, \delta)}$ and, for $i$ being equal either to $1$ or to $2$, we have ${\normalfont \mathrm{In}(\textbf{H}_i) = (\pi_i, \nu_i, \delta_i)}$.

\noindent (b) The following relations hold:
\begin{align*}
\pi &\geq \max(\pi_1-\nu_2,\pi_2-\nu_1),\\
\nu &\geq \max(\nu_1-\pi_2,\nu_2-\pi_1),\\
\pi &\leq \pi_1+\pi_2,\\
\nu &\leq \nu_1+\nu_2,\\
\forall i \in \llbracket 1,2\rrbracket,\; \pi+\nu+\delta &= \pi_i+\nu_i+\delta_i = n.
\end{align*} 
\end{lemma}
\noindent Full proof of Lemma \ref{lemma:H1plusH2} is provided in reference \cite{de1981inertia}. We now derive some boundary values for the inertia of $\bm{\gamma}^\Delta_\eta$.
\begin{proposition}\label{prop:boundaryEOM1}
Let {\normalfont{\textbf{X}}} and {\normalfont{\textbf{Y}}} be two $N\times (L-N)$ real matrices. Let $\bm{\gamma}^\Delta_\eta$ be the matrix defined as $\normalfont{\bm{\gamma}^\Delta_\eta}\coloneqq \normalfont{\textbf{A}}_x + \normalfont{\textbf{A}}_y,$ with $\normalfont{\textbf{A}}_x \coloneqq (-\normalfont{\textbf{XX}}\tmat) \oplus \normalfont{\textbf{X}}\tmat\normalfont{\textbf{X}}$ and $\normalfont{\textbf{A}}_y \coloneqq (-\normalfont{\textbf{YY}}\tmat) \oplus \normalfont{\textbf{Y}}\tmat\normalfont{\textbf{Y}}$. Let $(\pi_\eta, \nu_\eta, \delta_\eta)$ denote the inertia of ${\normalfont{\bm{\gamma}^\Delta_\eta}}$. Let $\pi_x$ denote $\pi({\normalfont{\textbf{X}}\tmat\normalfont{\textbf{X}}})$, $\nu_x$ denote $\nu({-\normalfont{\textbf{XX}}\tmat})$, $\pi_y$ denote $\pi({\normalfont{\textbf{Y}}\tmat\normalfont{\textbf{Y}}})$, and $\nu_y$ denote $\nu({-\normalfont{\textbf{YY}}\tmat})$. Then,
\begin{align*}
&|\pi_x - \pi_y|\leq \pi_\eta \leq \pi_x + \pi_y,\\
&|\nu_x - \nu_y\,|\leq \nu_\eta \leq \nu_x + \nu_y
\end{align*}
\end{proposition}
\begin{proof}
From Proposition \ref{prop:boundaryzeta} and its proof, adapted to the present matrices — i.e., $\textbf{A}_x$ and $\textbf{A}_y$, we immediately find that
\begin{equation}\label{eq:pieqnuEOM}
(\pi_x = \nu_x \leq \min(N,(L-N))) \quad \mathrm{and} \quad (\pi_y = \nu_y \leq \min(N,(L-N))).
\end{equation}
Applying Lemma \ref{lemma:H1plusH2} with $\textbf{H}_1 = \textbf{A}_x$ and $\textbf{H}_2 = \textbf{A}_y$ gives
\begin{align}
\label{eq:THREE}&\max(\pi_x - \nu_y,\pi_y-\nu_x) \leq \pi_\eta \leq \pi_x + \pi_y,\\
\label{eq:FOUR} &\max(\nu_x - \pi_y,\nu_y-\pi_x) \leq \nu_\eta \leq \nu_x + \nu_y.
\end{align}
Using $(\pi_x = \nu_x)$ and $(\pi_y = \nu_y)$ from \eqref{eq:pieqnuEOM} we get the desired result.
\end{proof}
\begin{proposition}\label{prop:boundaryEOM2}
Let $\pi_\eta$ and $\nu_\eta$ be the two natural integers defined in Proposition {\normalfont\ref{prop:boundaryEOM1}}. We simultaneously have 
\begin{align}
\label{eq:etaresult1}\pi_\eta &\leq \min(2N,(L-N)),\\
\label{eq:etaresult2}\nu_\eta &\leq \min (N,2(L-N)).
\end{align}
\end{proposition}
\begin{proof}
From the definition of positive- and negative semidefiniteness, we immediately find that
\begin{align}
\label{eq:simeq1} ((-\textbf{XX}\tmat - \textbf{YY}\tmat) \preceq 0) \land ((\textbf{X}\tmat\textbf{X} + \textbf{Y}\tmat\textbf{Y}) \succeq 0) \, &\Longrightarrow \, \pi_\eta \leq (L-N),\\
\label{eq:simeq2} ((-\textbf{XX}\tmat - \textbf{YY}\tmat) \preceq 0)\land ((\textbf{X}\tmat\textbf{X} + \textbf{Y}\tmat\textbf{Y}) \succeq 0) \, &\Longrightarrow \, \nu_\eta \leq N.
\end{align}
We need to prove that the proposition holds in three following cases:

$\;$

\noindent  \textit{Case 1}: $\min(N,(L-N))=N$ — From \eqref{eq:pieqnuEOM}, we have $\pi_x+\pi_y \leq 2N$. According to \eqref{eq:THREE}, we then have $\pi_\eta \leq 2N$. Combined with \eqref{eq:simeq1}, we get the desired result for $\pi_\eta$. Moreover, from \eqref{eq:simeq2} we know that $\nu_\eta \leq N$. Since we are in \textit{Case 1}, i.e., $N<(L-N)$, we also have $N<2(L-N)$, hence the desired result for $\nu_\eta$.

$\;$

\noindent \textit{Case 2}: $\min(N,(L-N))=(L-N)$ — From \eqref{eq:pieqnuEOM}, we have $\nu_x+\nu_y \leq 2(L-N)$. Thanks to \eqref{eq:FOUR} we then get $\nu_\eta \leq 2(L-N)$. Combined with \eqref{eq:simeq2}, we get \eqref{eq:etaresult2} again. Moreover, from \eqref{eq:simeq1} we know that $\pi_\eta \leq (L-N)$. Since we are in \textit{Case 2}, i.e. $(L-N)<N$, we also have $(L-N)<2N$, hence \eqref{eq:etaresult1} is also true here.

$\;$

\noindent \textit{Case 3}: $N=(L-N)$. Results \eqref{eq:etaresult1} and \eqref{eq:etaresult2}, which are both true in \textit{Case 1} and in \textit{Case 2}, are therefore true in \textit{Case 3}.
\end{proof}
\noindent These results should be compared with the results about the relaxed EOM--TDDFT \cite{etienne_boundary}, for which we only know that 
\begin{align*}
\pi_\eta &\leq \pi_\eta^{rlx}\leq L-N,\\
\nu_\eta &\leq \nu_\eta^{rlx}\leq N.
\end{align*}

\subsection{About exact and approximate transition pictures in CIS and beyond}

\subsubsection{About approximate transition pictures involving NDO's}

We deduce from the expression of the detachment and attachment density matrices that
\begin{proposition}\label{prop:NDOsnonpairing}
The natural difference orbitals pairing property is not universal.
\end{proposition}
\begin{proof} Unlike the CIS method for which the negative semidefinite and positive semidefinite blocks share $\mathrm{min}(N,(L-N))$ nonnegative eigenvalues according to Lemma \ref{theo:eigenvalues}, in the case of EOM--TDDFRT — which readily provides a counterexample to the NDO's pairing property — we do not have the same configuration since we have two sums of products of the type that was encountered in the negative semidefinite and positive semidefinite blocks in CIS, which means here that the $\eta$--type detachment/attachment 1--RDM's might not share eigenvalues — so the natural difference orbitals are not necessarily paired.
\end{proof}
\noindent We therefore lose the simplified departure-arrival orbital picture we had in the CIS case. We also find that the integral of the detachment/attachment density satisfies 
\begin{equation*}
\vartheta _\eta \geq 1.
\end{equation*}

\subsubsection{About approximate transition pictures involving NTO's}

\noindent We would like to discuss a proposition whose consequences are often met in the literature — that, given a transition, $\bm{\gamma}^\mathrm{T}$ contains the ``singles'' coefficients of the ansatz of the arrival state. For this sake, we need to set up a certain framework which, as we will show, allows to compare two situations — an auxiliary wave function is constructed from $\psi_0$ and the 1--TDM (among other things)  as a starting point, denoted below by $\mathcal{S}_2$ and $\mathcal{S}_3$ — with another one in which the 1--TDM is not the starting point of the construction, denoted below by $\mathcal{S}_1$.

\paragraph{{\thesubsubsection}.A Setting up a framework for the discussion} $\;$ \\

\noindent We are still interested in an $N$--electron system, described using $\mathcal{B}$ — whose composition is given in \eqref{eq:CanonicalBasis}. 
Let $\psi_0$ denote a one-determinant state built using exclusively the $N$ first elements of $\mathcal{B}$:
$$\psi_0 = \bigwedge_{i=1}^N \varphi _i.$$
In this framework, $\bm{\gamma}^\mathrm{T}_{0\rightarrow 1}$ will denote the 1--TDM between $\psi_0$ and a state denoted by $\Psi_1$. In $\mathcal{S}_1$, the $\Psi _1$ state is constructed, while it can be fictious in $\mathcal{S}_2$ and $\mathcal{S}_3$ — \textit{vide infra}. 

Let $S^{(0)}$ be the real linear span of $\{\psi_0\}$. Again, $q$ will be used here as the number equal to $\mathrm{min}(N,(L-N))$. For every $\ell$ in $\llbracket 1, q\rrbracket$ we define
$$S^{(\ell)} \coloneqq \mathrm{span}_{\mathbb{R}}Q^{(\ell)} = \left\lbrace \sum _{k=1}^{m(\ell)}\lambda_k \psi^{(\ell)} _k \, : \, \forall k \in \llbracket 1,m(\ell)\rrbracket, \, (\lambda_k \in \mathbb{R})\land (\psi^{(\ell)} _k \in Q^{(\ell)})\right\rbrace$$
where
\begin{eqnarray*} m(\ell) =
\left(\!\!\begin{array}{c}
N \\ \ell
\end{array}\!\!\right)
\left(\!\!\begin{array}{c}
L-N \\ \ell
\end{array}\!\!\right),
\end{eqnarray*}
and $Q^{(\ell)}$ is the set of all linearly-independent $N$--electron one-determinant states whose composition, made exclusively of elements of $\mathcal{B}$ — whose composition is given in \eqref{eq:CanonicalBasis} —, differs from the one of $\psi_0$ by exactly $\ell$ spinorbitals, i.e.,
\begin{align}\label{eq:Qell}
\forall \ell \in \llbracket 1, q\rrbracket,\; Q^{(\ell)} \coloneqq &\left\lbrace \bigwedge_{j=1}^N \varphi _{i_j} \; : \; (1\leq i_1 < \cdots < i_N \leq L)\land \left(\sum_{k=1}^N \left(1-\sum_{j=1}^N \braket{\varphi_k|\varphi_{i_j}}\right) = \ell\right)\right\rbrace,
\end{align}
with $\mathrm{Card \, Q^{(\ell)} = m(\ell)}$. The first condition in the definition of each $Q^{(\ell)}$ ensures that all the elements of $Q^{(\ell)}$ are linearly independent, so that a family built with exactly the elements of $Q^{(\ell)}$ is a suitable basis for the corresponding $S^{(\ell)}$ vector space. We also build
$$\texttt{S} \coloneqq \displaystyle\bigoplus_{\ell\in \llbracket 0,q\rrbracket} S^{(\ell)}.$$
Elements of $S^{(1)}$ are linear combinations of singly-excited determinant states; elements of $S^{(2)}$ are linear combinations of doubly-excited determinant states, etc. We also define $ (\hat{T}_1^{(\ell)})_{\ell\in\llbracket 1,q\rrbracket}$ and $(\hat{T}_2^{(\ell)})_{\ell\in\llbracket 1,q\rrbracket}$, two families of maps such that
\begin{equation}\label{eq:Tjmaps}
\forall j\in\llbracket 1,2 \rrbracket,\, \forall \ell \in \llbracket 1,q\rrbracket,\, \hat{T}_j^{(\ell)} \; : \; S^{(0)} \longrightarrow S^{(\ell)}.
\end{equation}
We also introduce ``zero'' maps: for every $\ell$ in $\llbracket 1,q\rrbracket$, the $\hat{Z}_{0\rightarrow \ell}$ is defined as
\begin{align*}
\hat{Z}_{0\rightarrow \ell}  \; : \; S^{(0)} &\longrightarrow S^{(\ell)} \\ \Phi &\longmapsto \hat{Z}_{0\rightarrow \ell}\Phi = \sum _{k=1}^{m(\ell)} 0 \psi^{(\ell)}_k.
\end{align*}
Let $\Psi_1$ and $\Psi _2$ be two all-excitation CI auxiliary wave functions built in $\texttt{S}$ according to
\begin{equation}\tag{\textsc{Rule} I}\label{eq:DefPsi1Psi2}
\forall j \in \llbracket 1,2\rrbracket, \, \Psi_j \coloneqq \sum _{\ell=1}^{q} \hat{T}^{(\ell)}_j\, \psi_0,
\end{equation}
with the $\hat{T}$ maps denoting the single-excitation level $(\ell=1)$, double-excitation level $(\ell=2)$, etc. We also build the matrix representation of the two single-excitation level maps in $\mathcal{B}$:
$$\forall j\in \llbracket 1, 2 \rrbracket, \, \textbf{T}_j^{(1)} \coloneqq \mathcal{M}(\hat{T}^{(1)}_j\!,\mathcal{B}).$$
$\hat{T}_1^{(1)}$ and $\hat{T}_2^{(1)}$ are the two linear combinations of single-excitation maps to be accounted for in the $(\psi _0 \rightarrow \Psi_1)$ and the $(\psi _0 \rightarrow \Psi_2)$ transitions:
$$\forall j \in \llbracket 1, 2 \rrbracket, \, \hat{T}_j^{(1)} = \sum _{i=1}^N \sum _{a=N+1}^L (\textbf{T}_j^{(1)})_{a,i}\,\hat{a}^\dag \hat{i}.$$
Due to their nature — see \eqref{eq:Tjmaps} —, $\textbf{T}_1^{(1)}$ and $\textbf{T}_2^{(1)}$, which are both belonging to $\mathbb{R}^{L\times L}$, are necessarily both block-diagonal, with solely their South-West block — i.e., the ``$vo$'' block — being susceptible of containing non-zero elements. Therefore, we have
\begin{equation}\label{eq:votoC2}
\forall j \in \llbracket 1, 2\rrbracket, \, \sum _{i=1}^N \sum _{a=N+1}^L (\textbf{T}_j^{(1)})_{a,i}\,\hat{a}^\dag \hat{i} = \sum _{r=1}^L \sum _{s=1}^L (\textbf{T}_j^{(1)})_{r,s}\,\hat{r}^\dag \hat{s}.
\end{equation}
Notice that ``all-excitation CI'' in this context is not understood as meaning that ``all the maps in \eqref{eq:DefPsi1Psi2} must be different from the zero map''. We rather mean that they \textit{can} be — in fact, at least one map must be different from the zero map for the obtained state to be different from $0\psi_0$. The expression ``all-excitation CI'' is also used rather than ``full CI'' because there is no scalar-multiple of $\psi_0$ appearing as a term in $\Psi_1$ or $\Psi_2$.

We define the 1--TDM's corresponding to the virtual $(\psi_0\longrightarrow\Psi_1)$ and $(\psi_0 \longrightarrow \Psi_2)$ transitions:
\begin{equation}\tag{\textsc{Rule} II}
\forall j\in \llbracket 1,2\rrbracket, \, \forall (r,s)\in C^2,\, (\bm{\gamma}^\mathrm{T}_{0\rightarrow j})_{s,r} \coloneqq \braket{\psi_0 | \hat{r}^\dag \hat{s}|\Psi_j}_\texttt{S}.
\label{eq:ruleI}
\end{equation}
We also consider the possibility to constrain $\textbf{T}_2^{(1)}$ to be such that
\begin{equation} \tag{\textsc{Rule} III}
\forall (i,a) \in I\times A, \,(\textbf{T}_2^{(1)})_{a,i} = (\bm{\gamma}^\mathrm{T}_{0\rightarrow 1})_{a,i}.
\label{eq:ruleII}
\end{equation}
We finally consider a third auxiliary wave function
\begin{equation} \tag{\textsc{Rule} IV}\label{eq:Psi3}
\Psi_3 \coloneqq \hat{\mathcal{T}}^{(1)}_3\, \psi_0 + \sum_{\ell=2}^q \hat{T}^{(\ell)}_3\,\psi _0,
\end{equation}
where $\hat{\mathcal{T}}^{(1)}_3$ is defined as
\begin{align*}
\hat{\mathcal{T}}^{(1)}_3 \; : \;\; S^{(0)} &\longrightarrow S^{(0)} \oplus S^{(1)}\\
\Phi &\longmapsto \sum_{r=1}^L\sum_{s=1}^L(\bm{\gamma}^\mathrm{T}_{0\rightarrow 1})_{r,s}\,\hat{r}^\dag\hat{s} \,\Phi,
\end{align*}
and the remaining maps used in the definition of $\Psi_3$ are simply built with the constraint that $(\forall \ell \in \llbracket 2,q\rrbracket,\, \hat{T}_3^{(\ell)} \; : \; S^{(0)} \longrightarrow S^{(\ell)})$. We finally define   $\bm{\gamma}^\mathrm{T}_{0\rightarrow 3}$ such that
\begin{equation} \tag{\textsc{Rule} {V}}\label{eq:GammaT03}
\forall (r,s)\in C^2,\, (\bm{\gamma}^\mathrm{T}_{0\rightarrow 3})_{s,r} \coloneqq \braket{\psi_0 | \hat{r}^\dag \hat{s}|\Psi_3}_\texttt{S}.
\end{equation}
and $\bm{\gamma}_0$ such that
\begin{align}
\forall (r,s)\in C^2,\, (\bm{\gamma}_{0})_{s,r} &\coloneqq \braket{\psi_0 | \hat{r}^\dag \hat{s}|\psi_0}_\texttt{S}\nonumber \\
&=  (\textbf{I}_o \oplus \textbf{0}_v)_{s,r}\tag{\textsc{Rule} VI}\label{eq:Gamma0}
\end{align}
where $\textbf{I}_o$ is the $N\times N$ identity matrix. On our way we have defined six construction rules: \\ $\;$ \\ — \ref{eq:DefPsi1Psi2} for constructing $\Psi _1$ and $\Psi _2$; \\ — \ref{eq:ruleI} for constructing $\bm{\gamma}^\mathrm{T}_{0\rightarrow 1}$ and $\bm{\gamma}^\mathrm{T}_{0\rightarrow 2}$; \\ — \ref{eq:ruleII} for a constrained construction of $\Psi_2$; \\ — \ref{eq:Psi3} for constructing $\Psi_3$;\\ — \ref{eq:GammaT03} for constructing $\bm{\gamma}^\mathrm{T}_{0\rightarrow 3}$; \\ — \ref{eq:Gamma0} for constructing $\bm{\gamma}_0$. \\ $\;$

\noindent Consider now the three following schemes:
\begin{align}
\nonumber{\mathcal{S}_1} \; : \; \psi_0 \,{\xrightarrow{\; (\mathrm{I}) \;}}\,  (\psi_0,\Psi_1) \,{\xrightarrow{\; (\mathrm{II}) \;}}\, &\bm{\gamma}^\mathrm{T}_{0\rightarrow 1} {\xrightarrow{\; (\mathrm{I})\,\mathrm{and}\,(\mathrm{III}) \;}} \, (\psi _0, \Psi _2) \, {\xrightarrow{\; (\mathrm{II}) \;}} \, \bm{\gamma}^\mathrm{T}_{0\rightarrow 2}, \\
\nonumber{\mathcal{S}_2} \; : \;\, &\bm{\gamma}^\mathrm{T}_{0\rightarrow 1} {\xrightarrow{\; (\mathrm{I})\,\mathrm{and}\,(\mathrm{III}) \;} } \, (\psi _0, \Psi _2) \, {\xrightarrow{\; (\mathrm{II}) \;} } \, \bm{\gamma}^\mathrm{T}_{0\rightarrow 2},\\
\nonumber{\mathcal{S}_3} \; : \;\,  &\bm{\gamma}^\mathrm{T}_{0\rightarrow 1} {\xrightarrow{\; (\mathrm{IV})\textcolor{white}{\,\mathrm{and}\,(\mathrm{II})}} } \, (\psi _0, \Psi _3) \, {\xrightarrow{\; (\mathrm{V}) \;} } \, \bm{\gamma}^\mathrm{T}_{0\rightarrow 3}.
\end{align}
The main difference between $\mathcal{S}_1$ and the two other schemes is that in the case of $\mathcal{S}_1$ we ensure that $\bm{\gamma}^\mathrm{T}_{0\rightarrow 1}$ is built from a one-determinant state and an all-excitation CI auxiliary wave function, while this condition is not present in $\mathcal{S}_2$ nor in $\mathcal{S}_3$. It is important to notice that in \ref{eq:Psi3}, $\mathcal{S}_2$, and $\mathcal{S}_3$, the ``1'' in $\bm{\gamma}^\mathrm{T}_{0\rightarrow 1}$ can point to a fictious state — that can be the true but unknowable state of interest for example. Notice also that, in $\mathcal{S}_3$, due to the nature of $\hat{\mathcal{T}}^{(1)}_3$, the $\Psi_3$ wave function is not an ``all-excitation CI''. We can now assert, about $\mathcal{S}_1$, that
\begin{lemma}\label{prop:S1-1}
Independently from the value of the coefficient-matrix elements in each $\hat{T}_1^{(\ell)}$ with $\ell$ belonging to $\llbracket 2,q\rrbracket$, we have that $\normalfont{\textbf{T}}_1^{(1)}$ is equal to $\bm{\gamma}^\mathrm{T}_{0\rightarrow 1}$ in $\mathcal{S}_1$.
\end{lemma}
\begin{proof}
In the appendices to reference \cite{etienne_comprehensive_2021}, we construct and prove a corollary to the fermionic, time-independent Wick theorem: 
\begin{corollary}\label{cor:wick}
The single-reference expectation value of a chain of $2K$ fermionic second quantization operators can be decomposed into a sum of $(2K-1)!!$ products of $K$ two-operator expectation values relatively to that reference, each product being affected by a sign corresponding to a permutation signature.
\end{corollary}
\noindent In Corollary \ref{cor:wick} above, $K$ is a non-zero natural integer, and $(2K-1)!!$ is the double factorial of $(2K-1)$ and should not be confused with $((2K-1)!)!$, which is simply the factorial of $(2K-1)!$.

For every value of $\ell$ between 2 and $q$, we construct the corresponding term in $\Psi_1$ by noticing that $\hat{T}^{(\ell)}_1 \, \psi _0$ is a linear combination of determinant states resulting from the action of $2\ell$ second quantization operators on $\psi_0$ such that no pair of these operators has a non-zero expectation value relatively to $\psi_0$. 

Let $\hat{Q}$ be any of these $2\ell$--dimensional second quantization operators chains. According to Corollary \ref{cor:wick}, for any $(r,s)$ couple in $C^2$, $\braket{\psi_0 | \hat{r}^\dag\hat{s}\hat{Q}|\psi_0}$ is equal to zero. Indeed, this expectation value can be decomposed into a sum of $(2\ell +1)!!$ products of $(\ell +1)$ two-operator expectation values relatively to $\psi_0$. Since $\ell$ belongs to $\llbracket 2, q\rrbracket$, for every term of the sum in Corollary \ref{cor:wick} there will be at least one pair of second quantization operators whose expectation value relatively to $\psi _0$ is zero. As a consequence, using \eqref{eq:votoC2} and Corollary \ref{cor:wick},
\begin{align*}
\forall (r,s) \in C^2, \, \braket{\psi_0|\hat{r}^\dag\hat{s}|\Psi_1} &= \braket{\psi_0|\hat{r}^\dag\hat{s}\hat{T}^{(1)}_1|\psi_0}\\
&= (\textbf{T}_1^{(1)})_{s,r}.
\end{align*}
Comparing this result with \ref{eq:ruleI} completes the proof.
\end{proof}
\noindent Notice that an alternative proof is possible and would imply the Slater-Condon rule for matrix elements of a one-body operator between two determinant states differing by at least two spinorbitals.
\begin{lemma}\label{prop:S12-T2}
Independently from the value of the coefficient-matrix elements in each $\hat{T}_2^{(\ell)}$ with $\ell$ belonging to $\llbracket 2,q\rrbracket$, we have that $\normalfont{\textbf{T}}_2^{(1)}$ is equal to $\bm{\gamma}^\mathrm{T}_{0\rightarrow 2}$ in $\mathcal{S}_1$.
\end{lemma}
\begin{proof}
Applying the proof strategy used for Lemma \ref{prop:S1-1} but slightly adapted in order to correspond to the $(\psi _0 \longrightarrow \Psi_2)$ transition rather than the $(\psi _0 \longrightarrow \Psi_1)$ one — i.e., by simply replacing ``1'' by ``2'' as the index to ``$\hat{T}$'', ``$\Psi$'', and ``$\textbf{T}$'' —, readily provides the desired result.
\end{proof}
\noindent A proposition can now be proved about elements of scheme $\mathcal{S}_1$, that will be of importance when comparing scheme $\mathcal{S}_1$ with scheme $\mathcal{S}_2$ and scheme $\mathcal{S}_3$:
\begin{proposition}\label{prop:S1-2}
Independently from the value of the coefficient-matrix elements in each $\hat{T}_2^{(\ell)}$ with $\ell$ belonging to $\llbracket 2,q\rrbracket$, we have that $\bm{\gamma}^\mathrm{T}_{0\rightarrow 1}$ is equal to $\bm{\gamma}^\mathrm{T}_{0\rightarrow 2}$ in $\mathcal{S}_1$.
\end{proposition}
\begin{proof}
Using Lemma \ref{prop:S12-T2} together with \ref{eq:ruleII} we immediately get the desired result.
\end{proof}
\noindent The next proposition tells us that under stronger constraints than in Proposition \ref{prop:S1-2} we can say more about elements of the corresponding, constrained $\mathcal{S}_1$ than in Proposition \ref{prop:S1-2}:
\begin{proposition}\label{prop:S1-3}
If $\Psi_1$ is a $\zeta$-state, and $\Psi_2$ is constructed using $\mathcal{S}_1$ and is also a $\zeta$-state, then $\Psi_1$ is equal to $\Psi _2$ and $\bm{\gamma}^\mathrm{T}_{0\rightarrow 1}$ is equal to $\bm{\gamma}^\mathrm{T}_{0\rightarrow 2}$:
$$\forall \Psi _1 \in W^\zeta, \, (\forall \ell \in \llbracket 2, q\rrbracket, \, \hat{T}^{(\ell)}_2 = \hat{Z}_{0\rightarrow \ell})\Longrightarrow [(\Psi_1 = \Psi _2) \land (\bm{\gamma}^\mathrm{T}_{0\rightarrow 1} = \bm{\gamma}^\mathrm{T}_{0\rightarrow 2})].$$
\end{proposition}
\begin{proof}
We already know from Proposition \ref{prop:S1-2} that, independently from the conditions of the present proposition,
$$\bm{\gamma}^\mathrm{T}_{0\rightarrow 1} = \bm{\gamma}^\mathrm{T}_{0\rightarrow 2}.$$ 
Moreover, according to Lemma \ref{prop:S1-1} and \ref{eq:ruleII}, we find that 
$$\textbf{T}^{(1)}_1 = \textbf{T}_1^{(2)}.$$ 
Therefore, applying the construction rule for $\Psi_2$ we explicitly wrote as \ref{eq:DefPsi1Psi2}, i.e.,
$$\Psi_2 = \sum _{i=1}^N \sum _{a=N+1}^L (\textbf{T}_1^{(1)})_{a,i}\, \hat{a}^\dag\hat{i} \, \psi _0 + \sum _{\ell = 2}^q \hat{T}^{(\ell)}_2 \, \psi _0,$$
we obtain, applying the constraint mentioned in the proposition, the following result:
\begin{align*}
(\forall \ell \in \llbracket 2, q\rrbracket, \, \hat{T}^{(\ell)}_2 = \hat{Z}_{0\rightarrow \ell}) \Longrightarrow \Psi _2 &= \sum _{i=1}^N \sum _{a=N+1}^L (\textbf{T}_1^{(1)})_{a,i}\, \hat{a}^\dag\hat{i} \, \psi _0.
\end{align*}
On the other hand, according to an other condition of the proposition, we have
$$\Psi_1 \in W^\zeta \Longrightarrow \forall \ell' \in \llbracket 2, q\rrbracket, \, \hat{T}^{(\ell')}_1 = \hat{Z}_{0\rightarrow \ell'}.$$
As a consequence, the expression of $\Psi_1$ simply reduces to
$$\Psi _1 = \sum _{i=1}^N \sum _{a=N+1}^L (\textbf{T}_1^{(1)})_{a,i}\, \hat{a}^\dag\hat{i} \, \psi _0.$$
Noticing that the final expression of $\Psi_1$ is the same as the final expression of $\Psi _2$ then completes the proof.
\end{proof}
\begin{proposition}\label{prop:S2variant}
Independently from the value of the coefficient-matrix elements in each $\hat{T}_2^{(\ell)}$ with $\ell$ belonging to $\llbracket 2,q\rrbracket$, we have that $\bm{\gamma}^\mathrm{T}_{0\rightarrow 1}$ is not necessarily equal to $\bm{\gamma}^\mathrm{T}_{0\rightarrow 2}$ in $\mathcal{S}_2$.
\end{proposition}
\begin{proof}
We will use a simple counterexample in order to prove this proposition: consider scheme $\mathcal{S}_2$. Its starting point is $\bm{\gamma}^\mathrm{T}_{0\rightarrow 1}$ without the constraint required in $\mathcal{S}_1$ that $\bm{\gamma}^\mathrm{T}_{0\rightarrow 1}$ is derived, using \ref{eq:DefPsi1Psi2} then \ref{eq:ruleI}, from a one-determinant state and an all-excitation CI relatively to that determinant state. Therefore, we can chose $\bm{\gamma}^\mathrm{T}_\eta$ from \eqref{eq:EOM-TDM} as our $\bm{\gamma}^\mathrm{T}_{0\rightarrow 1}$. Applying \ref{eq:ruleII} then gives
\begin{align*}
\Psi_2 =\sum _{i=1}^N\sum_{a=N+1}^L \left(\textbf{X}^\top\right)_{a-N,i} \, \hat{a}^\dag \hat{i}\, \psi _0  + \sum _{\ell=2}^q \hat{T}^{(\ell)}_2 \, \psi_0.
\end{align*}
Accordingly, we see that the one-body reduced transition density matrix we obtain using $\psi_0$, $\Psi _2$ and \ref{eq:ruleI} is
\begin{eqnarray*} 
\bm{\gamma}^\mathrm{T}_{0\rightarrow 2} =  \left( 
 \begin{array}{cc}
\textbf{0}_o  & \textbf{0}_{ov}   \\
\textbf{X}\tmat\!  &  \textbf{0}_v \\
 \end{array}\right)
\end{eqnarray*}
which is different from $\bm{\gamma}^\mathrm{T}_{0\rightarrow 1}$.
\end{proof}
\noindent Such a conclusion about $\Psi_2$ and the relation between $\bm{\gamma}^\mathrm{T}_{0\rightarrow 1}$ and $\bm{\gamma}^\mathrm{T}_{0\rightarrow 2}$ could be expected: due to \ref{eq:ruleII} and the structure of $\bm{\gamma}^\mathrm{T}_\eta$, some information is necessarily lost when constructing $\Psi_2$, that cannot be rebuilt when deriving $\bm{\gamma}^\mathrm{T}_{0\rightarrow 2}$ from $(\psi_0,\Psi_2)$. In a naively less expectable way, we can show that this is still true in $\mathcal{S}_3$, where — unlike in $\mathcal{S}_1$ and $\mathcal{S}_2$ — the full $\bm{\gamma}^\mathrm{T}_{0\rightarrow 1}$ is accounted for when building $\Psi_3$ from $\psi_0$ and $\bm{\gamma}^\mathrm{T}_{0\rightarrow 1}$:
\begin{proposition}\label{prop:S3variant}
Let the $\bm{\gamma}^\mathrm{T}_{0\rightarrow 1}$ matrix in $\mathcal{S}_3$ be partitioned as
\begin{eqnarray}\label{eq:partitionS3}
\normalfont \left(\begin{array}{cc}
\textbf{T}_{o\textcolor{white}{v}} & \textbf{T}_{ov}\\ \textbf{T}_{vo} & \textbf{T}_v
\end{array}\right) = \textbf{T}^{(1)}_{1,vo} + (\bm{\gamma}^\mathrm{T}_{0\rightarrow 1} - \textbf{T}^{(1)}_{1,vo})
\quad \normalfont{\textit{with}} \quad
\textbf{T}^{(1)}_{1,vo} \coloneqq \left(\begin{array}{cc}
\textbf{0}_{o\textcolor{white}{v}} & \textbf{0}_{ov}\\ \textbf{T}_{vo} & \textbf{0}_v
\end{array}\right),
\end{eqnarray}
\normalfont\textit{and with} $\textbf{T}_o \in \mathbb{R}^{N\times N}$. \textit{Then, still in $\mathcal{S}_3$,} $$\bm{\gamma}^\mathrm{T}_{0\rightarrow 3} = \mathrm{tr}(\textbf{T}_o)\bm{\gamma}_0 + \textbf{T}^{(1)}_{1,vo}\quad \normalfont{\textit{with}} \quad \bm{\gamma}_0 = (\textbf{I}_o \oplus \textbf{0}_v),$$ \textit{so that $\bm{\gamma}^\mathrm{T}_{0\rightarrow 3}$ is, in the general case, different from} $\bm{\gamma}^\mathrm{T}_{0\rightarrow 1}.$
\end{proposition}
\begin{proof}
We can decompose the four contributions of $\hat{\mathcal{T}}_3^{(1)}\psi_0$ according to the partition in \eqref{eq:partitionS3}:
\begin{align*}
\hat{\mathcal{T}}_3^{(1)}\psi_0 &= \sum_{i=1}^N \sum_{j=1}^N (\textbf{T}_o)_{i,j}\,\hat{i}^\dag\hat{j} \, \psi _0 \tag{D1}\label{eq:D1} \\
&+ \sum_{i=1}^N \sum_{a=N+1}^{L} (\textbf{T}_{ov})_{i,a-N} \,\hat{i}^\dag\hat{a}\,\psi_0 \tag{D2}\label{eq:D2}\\
&+ \sum_{a=N+1}^L \sum_{i=1}^{N} (\textbf{T}_{vo})_{a-N,i} \,\hat{a}^\dag\hat{i}\,\psi_0 \tag{D3} \label{eq:D3}\\
&+ \sum_{a=N+1}^L \sum_{b=N+1}^{L} (\textbf{T}_{v})_{a-N,b-N} \,\hat{a}^\dag\hat{b}\,\psi_0 \tag{D4}.\label{eq:D4}
\end{align*}
We immediately see that the second term and the fourth term — i.e., \eqref{eq:D2} and \eqref{eq:D4} — are both equal to $0\psi_0$. We also know that for every $(i,j)$ in $I^2$, we have $\hat{i}^\dag\hat{j}\,\psi_0 = \delta_{i,j}\psi _0$, so the first term — i.e., \eqref{eq:D1} — is $\mathrm{tr}(\textbf{T}_o)\psi_0$. Finally, the third term — i.e. \eqref{eq:D3} — alternatively reads
$$\sum_{a=N+1}^L \sum_{i=1}^{N} (\textbf{T}_{vo})_{a-N,i} \,\hat{a}^\dag\hat{i}\,\psi_0 = \sum_{a=N+1}^L \sum_{i=1}^{N} (\textbf{T}^{(1)}_{1,vo})_{a,i} \,\hat{a}^\dag\hat{i}\,\psi_0.$$
Accordingly, we can recast the action of $\hat{\mathcal{T}}_3^{(1)}$ on the $\psi _0$ state as $\hat{\mathcal{T}}_3^{(1)} \psi_0 = (\hat{D}_1+\hat{D}_3)\psi_0,$ with $\hat{D}_1 = \mathrm{tr}(\textbf{T}_o)\hat{\mathds{1}}_{S^{(0)}}$ and
$$\hat{D}_3 = \sum_{a=N+1}^L \sum_{i=1}^{N} (\textbf{T}^{(1)}_{1,vo})_{a-N,i} \,\hat{a}^\dag\hat{i}.$$
Using \ref{eq:Psi3} together with
$$\forall \ell \in \llbracket 2,q\rrbracket, \, \forall (r,s)\in C^2,\, \braket{\psi_0 |\hat{r}^\dag\hat{s}\hat{T}_3^{(\ell)}|\psi_0} = 0,$$
we obtain, using Corollary \ref{cor:wick} and \ref{eq:Gamma0}, and being inspired by \eqref{eq:votoC2} that we can perfectly adapt to $\textbf{T}_{1,vo}^{(1)}$ due to its structure,
\begin{align*}
\forall (r,s)\in C^2,\, \braket{\psi_0 |\hat{r}^\dag\hat{s}|\Psi_3} &= \braket{\psi_0 |\hat{r}^\dag\hat{s}(\hat{D}_1+\hat{D}_3)|\psi_0}\\
&= \mathrm{tr}(\textbf{T}_o)(\bm{\gamma}_0)_{s,r} + (\textbf{T}_{1,vo}^{(1)})_{s,r}.
\end{align*}
Applying \ref{eq:GammaT03} completes the proof.
\end{proof}
\noindent  In conclusion, we learned from propositions \ref{prop:S1-2}, \ref{prop:S2variant} and \ref{prop:S3variant} that (i) the one-body reduced transition density matrix is invariant under the use of an auxiliary wave function built according to $\mathcal{S}_1$, and that (ii) the one-body reduced transition density matrix is not invariant under the use of an auxiliary wave function built according to $\mathcal{S}_2$ or $\mathcal{S}_3$. It may be interesting to reconsider the whole case when $\psi_0$ is not constrained to be a one-determinant state.

\paragraph{{\thesubsubsection}.B EOM--TDDFRT and $\mathcal{S}_1$, $\mathcal{S}_2$, and $\mathcal{S}_3$} $\;$\\

\noindent Inserting $\bm{\gamma}^\mathrm{T}_\eta$ as $\bm{\gamma}^{\mathrm{T}}_{0\rightarrow 1}$ in both $\mathcal{S}_2$ and $\mathcal{S}_3$ would lead to a $\Psi_3$ state in $\mathcal{S}_3$ that would be equal to $\Psi_2$ in $\mathcal{S}_2$. That manipulation would lead to a $\bm{\gamma}^{\mathrm{T}}_{0\rightarrow 3}$ in $\mathcal{S}_3$ that would be equal to $\bm{\gamma}^{\mathrm{T}}_{0\rightarrow 2}$ in $\mathcal{S}_2$. Both matrices — i.e., $\bm{\gamma}^{\mathrm{T}}_{0\rightarrow 2}$ and $\bm{\gamma}^{\mathrm{T}}_{0\rightarrow 3}$ — would be different from the initially input $\bm{\gamma}^\mathrm{T}_\eta$ matrix.

In the EOM--TDDFRT framework, in the absence of de-excitations, \textbf{Y} vanishes, the Frobenius norm of \textbf{X} becomes equal to unity, and the resulting representation of the electronic transition is exactly consistent with the one of a $\zeta$--ansatz. We call this the Tamm-Dancoff Approximation (TDA).

If a TDA calculation yields an $(N \times (L-N))$ amplitudes matrix — our $\textbf{X}$ matrix —, taking a $\textbf{T}_1^{(1)}$ matrix in $\mathcal{S}_1$ partitioned as 
\begin{eqnarray*} 
  \left( 
 \begin{array}{cc}
\textbf{0}_o  & \textbf{0}_{ov}   \\
\textbf{X}\tmat\!  &  \textbf{0}_v \\
 \end{array}\right)
\end{eqnarray*}
 will then lead to an expression for the $\Psi _1$ auxiliary state in $\mathcal{S}_1$ that will have the exact same structure as the one, obtained using $\bm{\gamma}^\mathrm{T}_\eta$ as $\bm{\gamma}^{\mathrm{T}}_{0\rightarrow 1}$, of $\Psi_2$ in $\mathcal{S}_2$ and $\Psi_3$ in $\mathcal{S}_3$ — the structure will be the same, but not the content. The difference between the two approaches — $\mathcal{S}_1$ on one side, $\mathcal{S}_2$ and $\mathcal{S}_3$ on the other side — is that no information is lost in the successive constructions of $\mathcal{S}_1$ and the auxiliary wave functions are normalized but willingly simple in their structure, whereas information is lost along the $\mathcal{S}_2$ and $\mathcal{S}_3$ schemes, which will produce non-normalized auxiliary functions when the $\textbf{Y}$ matrix in $\bm{\gamma}^\mathrm{T}_\eta$ is not the $(N\times (L-N))$ zero matrix. This information is willingly lost by construction in $\mathcal{S}_2$ — due to \ref{eq:ruleII} — while it is mechanically lost in $\mathcal{S}_3$, though the input matrix is originally fully accounted for through $\hat{\mathcal{T}}^{(1)}_3$ in $\mathcal{S}_3$. Moreover, if multiple TDA solutions are produced, the auxiliary ansatzes constructed in $\mathcal{S}_1$ for different TDA solutions will be mutually orthogonal, which is no longer true when multiple non-TDA solutions are input in $\mathcal{S}_2$ and $\mathcal{S}_3$ for using the construction rules for building $\Psi_2$ and $\Psi_3$, respectively.
 
 In the EOM-TDDFRT framework, in the general case, we lose the property, important in CIS and recovered in TDA, that NTO's and NDO's are identical — see Proposition \ref{prop:CIS-NTOsNDOs}. NTO's are always paired — this is a property of the singular value decomposition — but we have shown with Proposition \ref{prop:NDOsnonpairing} that $\eta$-type NDO's are, in general, not paired. For this reason — and others — it seems, at first sight, more reasonable to use the NTO's for an orbital picture of the transition, but in the next paragraphs we will try to discuss this question using the framework built above.
 
\paragraph{{\thesubsubsection}.C Picturing molecular electronic transitions using NTO's} $\;$ \\

\noindent An approximate representation of molecular electronic transitions picturing them as a directional composition of right-to-left NTOs single-electron promotions as in CIS/TDA — the right-singular vectors correspond to the ``hole'' while the left-singular vectors correspond to the ``electron'' of the hole/electron simplified representation \cite{etienne_towards_2021} — is often met in the litterature, beyond the EOM-TDDFRT case: The elements of the one-body reduced transition density matrix are seen as the single-excitation coefficients in the single-electron-replacement (truncation) of a complete expansion of the ``all-excitation CI'' type, and the NTOs used to present this expansion in a compact way: The picture depicted here consists in a collection of 
\begin{center}
$i^\mathrm{th}$ \textit{transition-hole} NTO $\xrightarrow{\;\lambda_i\;}$ $i^\mathrm{th}$ \textit{transition-electron} NTO
\end{center}
\noindent representations. This may invite the reader to represent the transition as a transformation from an initial state, $\psi_0$, to a final (auxiliary) state, $\psi$, that could, at least partially, be accounted for in second quantization using a map like
$$ \hat{\mathcal{T}}_{\psi_0\rightarrow \psi}^{(1)} = \sum _{i=1}^L \lambda _i \, \hat{l}^\dag_i \hat{r}_i^{\textcolor{white}{\dag}}$$ 
in which the $\lambda _i$ coefficients are the singular values corresponding to the $i^\mathrm{th}$ couple of left (``$l$'') and right (``$r$'') singular vectors of the 1--TDM of the problem at stake. In the expression above, we have used the $\hat{l}^\dag _i$ and $\hat{r}_i$ operators to signify that they are second quantization operators in the NTO's basis: $\hat{l}^\dag _i$ creates an electron in the $i^\mathrm{th}$ left-NTO and $\hat{r}_i$ annihilates an electron in the $i^\mathrm{th}$ right-NTO. At this point the connection with \eqref{eq:CISwf} is obvious: It is then much tempting, being culturally impregnated by the manipulation of the same objects in CIS — where these manipulations leave the description exact —, and consistently with the directional ``transition-hole NTO $\rightarrow$ transition-electron NTO'' picture, to represent the transition as taking place between $\psi_0$ and
$$\psi \simeq \sum _{i=1}^L \lambda _i \, \hat{l}^\dag_i \hat{r}_i^{\textcolor{white}{\dag}}\,\psi_0$$
with optionally some higher-order excitations that would not be accounted for in the NTO picture. Since the sum above runs over all the NTO's, we can write the $\hat{\mathcal{T}}_{\psi_0\rightarrow \psi}^{(1)}$ map, backtransformed in $\mathcal{B}$, as
$$\hat{\mathcal{T}}_{\psi_0\rightarrow \psi}^{(1)} = \sum _{r=1}^L\sum_{s=1}^L \left(\bm{\gamma}^\mathrm{T}_{\psi_0\rightarrow \psi} \right)_{r,s} \, \hat{r}^\dag \hat{s},$$
leading to the following expression for the (auxiliary) function $\psi$:
\begin{equation}\label{eq:krylov}
\psi =  \sum _{r=1}^L\sum_{s=1}^L \left(\bm{\gamma}^\mathrm{T}_{\psi_0 \rightarrow \psi} \right)_{r,s} \, \hat{r}^\dag \hat{s}\, \psi _0 \left(+ \cdots\right).
\end{equation}
where the three dots ``$\cdots$'' point to higher-order excitations. Such an expression is explicitly present (without parentheses around ``$+\cdots$'') in the literature. 

\noindent It seems to us of absolute seminal importance to notice that an expression like \eqref{eq:krylov} without extra-propositional information leaves undetermined whether $\psi$ is an auxiliary function built from a 1--TDM (and optionally other things) within scheme $\mathcal{S}_3$, or if $\psi$ is an optionally truncated or all-excitation CI function about which we say that we can identify, from a reference state $\psi_0$, the singles coefficient matrix with the $(\psi_0 \longrightarrow \psi)$ 1--TDM, which places the proposition in the extra-propositional context $\mathcal{S}_1$.
\label{par:C}

\paragraph{{\thesubsubsection}.D The problem of the NTO picture directionality} $\;$ \\

\noindent We believe that picturing molecular electronic transitions using a directional NTO representation deserves to be questioned. On our way discussing this picture we will use the abovementioned $(\mathcal{S}_1,\mathcal{S}_2,\mathcal{S}_3)$ framework. 

Before all we would like to discuss the terms ``occupied'' and ``virtual'' NTO's that are often met in the literature. This, again, may be a cultural heritage from the literature about the (1--TDM, NTO) framework applied to the CIS calculation method — e.g., the pioneering work of Luzanov and co-workers \cite{luzanov_application_1976} as well as the work of Mayer \cite{mayer_using_2007} and of Surján \cite{surjan_natural_2007}. When we are beyond CIS, looking at a directional NTO picture with ``occupied'' and ``virtual'' NTO's we immediately see the strong connection with \eqref{eq:CISwf}, but we think that such a picture in this context is questionable. EOM--TDDFRT is paradigmatic for questioning the extension of this convention beyond CIS: The square, non-symmetric $\eta$-type 1--TDM is susceptible of having non-zero entries in both the ``$vo$'' and the ``$ov$'' blocks. Therefore, its left- and right-singular vectors are both susceptible of having non-zero components in the occupied \textit{and} in the virtual canonical subspaces. For this reason we think that it may be misleading — unless the \textbf{y} components are all close to zero — to speak in terms of ``occupied'' and ``virtual'' NTO's outside the CIS/TDA framework, in which this vocabulary is perfectly adequate — \textit{occupied} (respectively, \textit{virtual}) CIS/TDA NTO's are obtained by a rotation of the occupied (respectively, virtual) canonical space. 

If a directional $\eta$-type NTO picture is put in correspondence with an auxiliary ansatz as in \eqref{eq:krylov}, that ansatz is of the type we meet in $\mathcal{S}_3$. Hence, according to Proposition \ref{prop:S3variant}, the 1--TDM that would be derived from such an ansatz would differ from the original-solution 1--TDM. Though each individual NTO has been constructed from the SVD of a matrix which fully accounts for the de-excitations through the $\textbf{Y}$ matrix in the ``$ov$'' block, using \textit{all of them at the same time} in a ``departure/arrival'' picture based on the $\mathcal{S}_3$ scheme leads the de-excitation contributions to fully disappear from the picture as a consequence of Proposition \ref{prop:S3variant}, so the information loss will make the one-body reduced transition density function and some transition properties such as the transition dipole moment corresponding to this auxiliary ansatz inconsistent with the original, native solution. This, we believe, has something to do with the fact that NTO's are not meant, in the general case, to be part of a ``departure/arrival'' picture understood in the sense that has been depicted in paragraph C right above — \textit{vide infra}, and see reference \cite{etienne_towards_2021} —, so that one should not even enter the $\mathcal{S}_3$ scheme at all for this precise reason. Actually, since using the auxiliary ansatz $\psi$ of \eqref{eq:krylov} in TDDFRT would lead to losing the de-excitation contributions, there has been a quest for finding physically sound auxiliary ansatzes — this is the so-called TDDFRT ``assignment problem'' — so that in practice one does not construct a $\psi$ auxiliary ansatz according to \eqref{eq:krylov} and avoids entering in the $\mathcal{S}_3$ scheme. Some of those explicit auxiliary ansatzes are built using \ref{eq:DefPsi1Psi2} restricted to the ($\ell =1$) level — hence, entering $\mathcal{S}_1$ with those auxiliary ansatzes is in line with the conditions of propositions \ref{prop:S1-2} and \ref{prop:S1-3}. These explicit auxiliary ansatzes are studied in the last part of this contribution. This problem of course does not occur in the case of CIS/TDA, but we do learn from this part of the paper that the coincidence between $\bm{\gamma}^\mathrm{T}_{0\rightarrow 1}$ and $\bm{\gamma}^\mathrm{T}_{0\rightarrow 2}$ (respectively, $\bm{\gamma}^\mathrm{T}_{0\rightarrow 3}$) in $\mathcal{S}_2$ (respectively, $\mathcal{S}_3$) when the input $\bm{\gamma}^\mathrm{T}_{0\rightarrow 1}$ matrix is the CIS/TDA one is purely accidental.

\subsubsection{About approximate transition pictures involving density (matrices)}

When the approximate picture of the transition is given in terms of one-body reduced density functions rather than one-body wavefunctions, the
\begin{center}
\textit{transition-hole} density $\longrightarrow$ \textit{transition-electron} density
\end{center}
representation can appear. Since the EOM--TDDFRT transition-hole ($h$) and the transition-electron ($e$) one-body reduced density matrices have the following structure \cite{etienne_towards_2021}
\begin{align*}
\bm{\gamma}_\eta^h &\coloneqq \textbf{XX}^\top \oplus \textbf{Y}\tmat\textbf{Y} = (\bm{\gamma}^\mathrm{T}_\eta)\tmat\bm{\gamma}^\mathrm{T}_\eta, \\
\bm{\gamma}_\eta^e &\coloneqq \textbf{YY}^\top \oplus \textbf{X}\tmat\textbf{X} = \bm{\gamma}^\mathrm{T}_\eta(\bm{\gamma}^\mathrm{T}_\eta)\tmat, 
\end{align*}
we see that, though the trace of $\bm{\gamma}^d_\eta$, $\bm{\gamma}_\eta ^a$, $\gamma_\eta ^h$, and $\gamma_\eta ^e$ are equal — and superior or equal to unity —, the detachment (respectively, attachment)  density matrix solely involves, with a contribution susceptible of being non-zero, the orbitals that are occupied (respectively, unoccupied) in the auxiliary ground state — i.e., the Kohn-Sham state —, while both the transition-hole and the transition-electron density matrices involve, with a contribution susceptible of being non-zero, orbitals from the full canonical space.

\subsubsection{General discussion}

We know that additional information in detachment/attachment — for example information that is coming, for a one-determinant reference state, from multiple excitations in the excited-state ansatz — cannot be retrieved in the 1--TDM. This information is also lost in the transition-hole and transition-electron density, or transition-hole NTO and transition-electron NTO pictures. There would probably be a true interpretative ambiguity when comparing the hole/electron and the detachment/attachment pictures if we could not suggest a disambiguation based on the fact that the hole/electron density matrices and the NTOs are not meant to bear the same physical information as the detachment/attachment density matrices and the NDOs: the (1--TDM, transition-hole/transition-electron, NTO) framework corresponds to \textit{coherences}, i.e. the coupling between stationary states that are not eigenstates of a field-perturbed Hamiltonian, and is used to build a one-hole/one-particle (sometimes abbreviated $1hp$, or simply $hp$) simplified, non-directional model of the transition — hence a ``$\leftrightarrow$'' symbol should be more appropriate than a ``$\rightarrow$'' symbol between the two NTO's of each couple in an NTO picture of the transition —, while the (1--DDM, detachment/attachment, NDO) framework corresponds to \textit{population transfers} and is conceived, through reduced quantities, as a framework for directional — i.e., ``departure/arrival'' — representation of the transition that does not obliterate some of the information about the system and the transition that may be absent from a $1hp$ model of the transition. However, as Proposition \ref{prop:univ} below shows, this directionality cannot be retrieved directly from the orbital picture of the (1--DDM, detachment/attachment, NDO) framework for visualization purposes: we actually close this section by a general proposition regarding the natural-orbital representation of molecular electronic transitions:
\begin{proposition} \label{prop:univ}
There is no universal, exact ``departure/arrival'' visualization picture that can be drawn from natural (transition or difference) orbitals in the general case.
\end{proposition}
\begin{proof}
Showing that NDO's pairing is not universal, and that in the general case — i.e., outside the CIS/TDA case —, NTO's may differ from the NDO's and are not meant to provide a ``departure/arrival'' picture of the transition, provides the requested proof.
\end{proof}
\noindent Without such a disambiguation, the representation of molecular electronic transitions would be either equivocal and arbitrary, or incomplete.

As a consequence of Proposition \ref{prop:univ} and of the discussion above, we conclude that among the objects discussed here only the detachment/attachment densities should be seen as being part of a ``departure/arrival'' simplified picture designed for visualization purposes.

Notice that, aside the ``state-coupling'' and the ``departure/arrival'' pictures,  a third characterization perspective is possible — the ``before/after'' picture — involving visualization and integration of the one-body reduced difference density function defined in \eqref{eq:1DDf}. This perspective is extensively studied in \cite{etienne_boundary}, together with its quantitative connexions with the detachment/attachment representation of molecular electronic transitions.

\subsection{Study of some explicit TDDFRT auxiliary many-body wavefunctions}
\noindent In this subsection we will discuss three alternative pictures to the original equation-of-motion TDDFRT, using auxiliary many-body wavefunctions for the electronic excited states.
\subsubsection{Four $\zeta$--type AMBW's}
\noindent We will introduce here four TDDFRT AMBW that share the properties that are common to the elements of $W^\zeta$, which were extensively studied in section \ref{sec:CIS}. The four ansatzes are defined using elements of \eqref{eq:casida}. 

The first of them, denoted $\Psi^{[1]}$ below, was originally proposed in Ref. \cite{casida_time-dependent_1995} and extensively used in the context of non-adiabatic dynamics \cite{crespo-otero_recent_2018,plasser_surface_2014,tapavicza_trajectory_2007,tavernelli_non-adiabatic_2009,tavernelli_nonadiabatic_2009,tavernelli_nonadiabatic_2009-1,tavernelli_nonadiabatic_2010}. That auxiliary ansatz reads
\begin{equation}\label{eq:tavernelliAMBW}
\Psi^{[1]} \coloneqq \sum _{i=1}^N\sum_{a=N+1}^L \delta_{\sigma_i,\sigma_a} \sqrt{\dfrac{\epsilon_a - \epsilon_i}{\omega}}(\textbf{z}_+)_{ia} \hat{a}^\dag\hat{i}\psi_0.
\end{equation} 
In \eqref{eq:tavernelliAMBW}, the ``$\epsilon$'' symbol denotes the Kohn-Sham energy of the corresponding spin-orbital, the ``$\sigma$'' symbol denotes the spin part of the spinorbital — see \eqref{eq:CanonicalBasis} —, the $\textbf{z}_+$ vector is defined as
$$\textbf{z}_+\coloneqq \sqrt{\omega}\left(\textbf{A}-\textbf{B}\right)^{-1/2}(\textbf{x}+\textbf{y}).$$
In reference \cite{hu2010nonadiabatic}, Hu and co-workers show the equivalence of two formulations of the TDDFRT non-adiabatic couplings and introduce another AMBW, denoted $\Psi^{[2]}$ below. They explain that the applicability of either $\Psi^{[1]}$ or $\Psi^{[2]}$ actually depends on the operator that is present in the inner product necessary for computing the quantity that is at stake — this is also explained in reference \cite{wang_nactddft_2021}. Their auxiliary ansatz reads
\begin{equation}\label{eq:HuAMBW}
\Psi^{[2]} \coloneqq \sum _{i=1}^N\sum_{a=N+1}^L \delta_{\sigma_i,\sigma_a} \sqrt{\dfrac{\omega}{\epsilon_a - \epsilon_i}}(\textbf{z}_+)_{ia} \hat{a}^\dag\hat{i}\psi_0.
\end{equation} 
The construction of $\Psi^{[1]}$ and $\Psi^{[2]}$ is conditioned to the fact that $(\textbf{A}-\textbf{B})$ is positive definite.

\noindent In reference \cite{luzanov_electron_2010}, Luzanov and Zhikol show that, in certain conditions detailed below, ``the procedure for reducing \eqref{eq:casida} to the Hermitian form is not unique''. They produce two families of auxiliary ansatzes, solution to two separate ``Hermitian problems'', each family being expected to be orthonormal if condition \eqref{eq:x2-y2} is fulfilled. The representative, generic ansatz for these two families of auxiliary states will be given below, denoted by $\Psi^{[3]}$ and $\Psi^{[4]}$. They are directly defined through their $\textbf{c}$ vector — see \eqref{eq:CISwf}: the $\textbf{c}^{[3]} = \textbf{z}_+$ (respectively, $\textbf{c}^{[4]} = \textbf{z}_-$) vector contains the components of $\Psi^{[3]}$ (respectively, $\Psi^{[4]}$) in the singly-excited determinant-state basis derived from $Q^{(1)}$ — see \eqref{eq:Qell}:
\begin{equation}\label{eq:zhikolAMBW1}
\Psi^{[3]} \coloneqq \sum _{i=1}^N\sum_{a=N+1}^L (\textbf{z}_+)_{ia} \hat{a}^\dag\hat{i}\psi_0,
\end{equation}
\begin{equation}\label{eq:zhikolAMBW2}
\Psi^{[4]} \coloneqq \sum _{i=1}^N\sum_{a=N+1}^L (\textbf{z}_-)_{ia} \hat{a}^\dag\hat{i}\psi_0. 
\end{equation}
In \eqref{eq:zhikolAMBW2}, we find the $\textbf{z}_-$ vector defined as
$$\textbf{z}_- \coloneqq\sqrt{\omega}(\textbf{A}+\textbf{B})^{-1/2}(\textbf{x}-\textbf{y}).$$
We see that the construction of $\Psi^{[3]}$ is conditioned to the fact that $(\textbf{A}-\textbf{B})$ is positive definite, and that the construction of $\Psi^{[4]}$ is conditioned to the fact that $(\textbf{A}+\textbf{B})$ is positive definite.

Since $\Psi^{[1]}$ is the most extensively used in calculation codes, we suggest to study it in more details, for some objects that can be derived from it are very close in their form, but different from EOM objects. After a necessary normalization, we will inspect the properties of the 1--TDM and 1--DDM one can derive from the $\Psi^{[1]}$ ansatz, and the structure of the related objects. The theoretical study and use of the $\Psi^{[1]}$ ansatz is made easy when considering pure-DFT exchange-correlation functionals, for in theses cases
\begin{equation*}
\forall (i,j)\in I^2,\forall (a,b)\in A^2,\, (\textbf{A}-\textbf{B})_{ia,jb} = (\epsilon _a - \epsilon _i)\delta_{i,j}\delta_{a,b}\delta_{\sigma_i,\sigma_a},
\end{equation*}
hence, $\Psi^{[1]}$ becomes
\begin{equation} \label{eq:CasidaAnsatzText}
\Psi^{[1]} \coloneqq \sum _{i=1}^N \sum _{a=N+1}^{L}  {\left(\textbf{x}_{ia}+\textbf{y}_{ia}\right)} \hat{a}^\dag\hat{i}\psi_0.
\end{equation} 
Notice that, in the same conditions, $\Psi^{[2]}$ reads
\begin{equation*}
\Psi^{[2]} \coloneqq \sum _{i=1}^N \sum _{a=N+1}^{L}  {\left(\textbf{x}_{ia}-\textbf{y}_{ia}\right)} \hat{a}^\dag\hat{i}\psi_0.
\end{equation*} 
When inspecting \eqref{eq:CasidaAnsatzText}, we see that multiplication of $\Psi^{[1]}$ by the inverse of 
\begin{equation}\label{eq:renormalization}
z_{[1]}^{1/2}   \coloneqq  \sqrt{(\textbf{x}+\textbf{y})\tmat  (\textbf{x}+\textbf{y}) } 
\end{equation}
is required to yield a normalized ansatz belonging with the corresponding 1--DDM having a zero trace, depicting a number-conserving electronic transition. Indeed, without the normalization, the norm of the excited-state wavefunction would be the square root of $z_{[1]}$ from (\ref{eq:renormalization}), and the trace of the 1--DDM would be $N \!\left({z}_{[1]} - 1\right)$, which might differ from zero. Notice that such a normalization requirement is, of course, also true for pure-DFT $\Psi^{[2]}$.

Using the same manipulations as in (\ref{eq:endrecastsquare}), the TDDFRT renormalized ansatz produces a zero-trace 1--DDM
\begin{equation*}
\textbf{W}_d \oplus \textbf{W}_a =  z_{[1]}  \, \bm{\gamma}^\Delta_{{[1]}} \in \mathbb{R}^{L \times L} 
\end{equation*} 
with
\begin{align*}
-\textbf{W}_d &\coloneqq  \left(\textbf{X}+\textbf{Y} \right)\left(\textbf{X}+\textbf{Y} \right)\!\tmat, \\
\textbf{W}_a &\coloneqq \left(\textbf{X}+\textbf{Y} \right)\!\tmat\!\left(\textbf{X}+\textbf{Y} \right).
\end{align*}
The 1--TDM, $\bm{\gamma} ^{\mathrm{T}}_{{[1]}}$, is nilpotent:
\begin{eqnarray*}
 \left( 
 \begin{array}{cc}
\textbf{0}_{o\textcolor{white}{v}}  & \textbf{X}+\textbf{Y}   \\
\textbf{0}_{vo}  &  \textbf{0}_v \\
 \end{array}  \right) =  z  _{[1]}  ^{1/2}\left(\bm{\gamma} ^{\mathrm{T}}_{{[1]}}\right)\tmat \in \mathbb{R}^{L \times L}.  
\end{eqnarray*}
According to Lemma \ref{lemma:prodPD}, we have that 
\begin{align*}
- \textcolor{black}{\left(\textbf{X}+ \textbf{Y} \right)\left(\textbf{X}+ \textbf{Y} \right)\!\tmat} &\preceq 0, \\
\textcolor{white}{--}\textcolor{black}{\left(\textbf{X}+ \textbf{Y} \right)\!\tmat\!\left(\textbf{X}+ \textbf{Y} \right)} &\succeq 0.  
\end{align*}
Therefore, the detachment (${\bm{\gamma} ^{d} _{{[1]} }}$) and attachment (${\bm{\gamma} ^{a} _{{[1]} }}$) 1--RDM's read
\begin{align*}
\textcolor{black}{\left(\textbf{X}+ \textbf{Y} \right)\left(\textbf{X}+ \textbf{Y} \right)\!\tmat} \oplus \textbf{0}_{\textcolor{black}{v}}&= z_{[1]} \,\textcolor{black}{\bm{\gamma} ^{d} _{{[1]} }}, \\
 \textbf{0}_{\textcolor{black}{o}} \oplus \textcolor{black}{\left(\textbf{X}+ \textbf{Y} \right)\!\tmat\!\left(\textbf{X}+ \textbf{Y} \right)} &= z_{[1]}  \,\textcolor{black}{\bm{\gamma} ^a _{{[1]}}} 
\end{align*}
and share eigenvalues — see Lemma \ref{theo:eigenvalues} and equations \eqref{eq:OTTO} and \eqref{eq:VTTV} — and can be derived without requiring the diagonalization of the 1--DDM. Their trace is equal to $1$.

The transition density matrix is nilpotent, and its non-zero elements can be collected into a rectangular $\textbf{T}^\top_{[1]}$ matrix that is such that 
\begin{equation*}
z _{[1]}  ^{-1/2}\left(\textbf{X}^\top+\textbf{Y}^\top\right) = \textbf{T}^\top_{[1]} \in \mathbb{R}^{(L-N) \times N}.  
\end{equation*}
This matrix can be diagonalized by an SVD to produce its \textcolor{black}{left} and \textcolor{black}{right} singular vectors, i.e., the \textit{left} and \textit{right} \textit{a}NTOs (\textit{auxiliary} Natural Transition Orbitals), eigenvectors of $\textcolor{black}{\bm{\gamma} ^{a} _{{[1]}}}$ and $\textcolor{black}{\bm{\gamma} ^{d} _{{[1]}}}$ paired by their eigenvalue. The one-body reduced transition density kernel \cite{etienne_towards_2021} for the $\Psi^{[1]}$ auxiliary ansatz reads
\begin{align*}
\gamma ^{\mathrm{T}}_{[1]} (\textbf{s}_1 ; \textbf{s}_1') 
&= z_{[1]}^{-1/2}\sum _{i=1}^N\sum _{a=N+1}^L (\textbf{x}+\textbf{y})_{ia} \, \varphi _a (\textbf{s}_1) \varphi _i (\textbf{s}_1')
\end{align*} 
where $(\textbf{s}_1) \coloneqq (\textbf{r}_1,s_1)$ and $(\textbf{s}_1') \coloneqq (\textbf{r}_1',s_1')$. For the EOM approach, it reads
\begin{align*}
\gamma ^{\mathrm{T}}_\eta (\textbf{s}_1 ; \textbf{s}_1') = \sum _{i=1}^N\sum _{a=N+1}^L \left.\textbf{x}_{ia} \, \varphi _a (\textbf{s}_1) \varphi _i (\textbf{s}_1')\right. \nonumber + \sum _{j=1}^N\sum _{b=N+1}^L\left.\textbf{y}_{jb} \, \varphi _j (\textbf{s}_1) \varphi _b (\textbf{s}_1')\right. .
\end{align*}
Though the two kernels are different, the corresponding one-body transition densities are identical, up to a scaling factor:
\begin{align*}\nonumber
z_{[1]}^{1/2}n^{\mathrm{T}}_{{[1]}}(\textbf{s}_1) &= n^{\mathrm{T}}_{\eta}(\textbf{s}_1) \\ &=  \sum _{i=1}^N\sum _{a=N+1}^L (\textbf{x}+\textbf{y})_{ia} \, \varphi _i (\textbf{s}_1) \varphi _a (\textbf{s}_1).
\end{align*}
In practice there are usually more than one excited state computed at the same time during a calculation. Hence, a set of data and matrices is produced as the outcome of such a calculation, and the type--$\Psi^{[1]}$ AMBW's one can construct based on the proposed scheme are not orthogonal to each other. After orthogonalizing a set of type--$\Psi^{[1]}$ AMBW's, we verify that our conclusions about the $\Psi^{[1]}$ ansatzes are transferrable when orthogonalization of the auxiliary excited states is done \textit{a posteriori}. For that sake we consider $M$ normalized but non-orthogonal type--$\Psi^{[1]}$ ansatzes. Any of these ansatzes reads
\begin{equation*}
{\Psi _m^{[1]}} = \sum _{i=1}^N \sum _{a=N+1}^L (\textbf{c}_m^{[1]})_{ia}\,\hat{a}^\dag\hat{i}{\psi _0},
\end{equation*}
with $(m \in S)$, and $z_{m}^{1/2}\,(\textbf{c}_m^{[1]})_{ia} = \left(\textbf{x}_m\right)_{ia}+\left(\textbf{y}_m\right)_{ia}$ — we have explicitly labeled the number of the excited state, $m$, using subscripts. For every $m$ in $S$, the $z_m$ number used for normalization is equal to $(\textbf{x}_m+\textbf{y}_m)\tmat(\textbf{x}_m+\textbf{y}_m)$. The overlap between the states is defined as
\begin{equation*}
\forall (m,n)\in S^2, \, ({\textbf{S}})_{m,n} \coloneqq \braket{\Psi _m ^{[1]} | \Psi _n ^{[1]}} = (\textbf{c}_m^{[1]})\tmat \textbf{c}_n^{[1]}.
\end{equation*}
One can orthonormalize the set of primitive AMBW's by considering a linear combination of them, for example using Löwdin's orthogonalization scheme:
\begin{align*}
\forall k \in S,\; \psi _k^\perp &\coloneqq \sum _{p=1}^M \left({\textbf{S}}^{-1/2}\right)_{p,k}{\Psi _p^{[1]}} \\&= \sum _{i=1}^N \sum _{a=N+1}^L \underbrace{\left(\sum _{p=1}^M\left({\textbf{S}}^{-1/2}\right)_{p,k} (\textbf{c}_p^{[1]})_{ia}\right)}_{\displaystyle(\textbf{c}_k^\perp)_{ia}}\hat{a}^\dag\hat{i}{\psi _0}
\end{align*}
which is equivalent to writing
\begin{equation}\label{eq:normalizedCasidaAnsatz}
\forall k \in S,\; \psi _k ^\perp = \sum _{i=1}^N \sum _{a=N+1}^L (\textbf{c}_k^\perp)_{ia}\,\hat{a}^\dag\hat{i}\psi _0.
\end{equation}
Let $W^\perp$ be the set of these $M$ orthonormal states. It comes that
\begin{equation*}
W^\perp \subset W^\zeta.
\end{equation*}
According to \eqref{eq:normalizedCasidaAnsatz} showing a normalized linear combination of singly-excited configurations, we see that the same rules for constructing the 1--DDM and 1--TDM corresponding to elements of $W^\zeta$ also apply to elements of $W^\perp$: A $\zeta$--type 1--TDM can be constructed for each element of $W^\perp$, and the non-zero matrix elements of that 1--TDM can be encoded into an ${(L-N)\times N}$ matrix leading to the derivation of projected-auxiliary NTO's. One can also obtain a zero-trace $\zeta$--type 1--DDM, and $\zeta$--type detachment/attachment 1--RDM's — with a trace equal to unity — from this picture. Note however that the representation of the individual electronic transitions will rely on the size and content of the set of excited states ansatzes used for constructing the abovementioned objects.
\subsubsection{The AMBW's of Luzanov and Zhikol}
Due to the non-uniqueness of the electronic-transition representation involving the $\textbf{z}_+$ or $\textbf{z}_-$ components into a nilpotent transition operator — \textit{vide supra} —, Luzanov and Zhikol have, in the same paper \cite{luzanov_electron_2010}, proposed an alternative, unique scheme for reducing the generalized eigenvalue problem \eqref{eq:casida} into a Hermitian eigenvalue problem: the \eqref{eq:casida} central equation, that we can rewrite as $\bm{\Delta} \ket{u} = \omega \textbf{J}\ket{u}$, with
\begin{eqnarray*}
\bm{\Delta} \coloneqq \left(\textcolor{black}{ \begin{array}{cc}
\textbf{A}^{\textcolor{white}{*}}  & \textbf{B}  \\ \textbf{B}^{\textcolor{white}{*}}  & \textbf{A} 
\end{array}}\right),\quad
\textbf{J} \coloneqq \left( \begin{array}{cc}
\textbf{I} & \textcolor{white}{-}\textbf{0} \\ \textbf{0} & -\textbf{I}
\end{array}\right),\quad
\ket{u} \coloneqq \left(\textcolor{black}{ \begin{array}{c}
\textbf{x}  \\ \textbf{y}
\end{array}}\right),
\end{eqnarray*}
is, upon the condition that both $\left(\textbf{A}+\textbf{B}\right)$ and $\left(\textbf{A}-\textbf{B}\right)$ are positive semidefinite, equivalent to a Hermitian eigenvalue problem whose eigenvectors are
\begin{align*}
\ket{u_\star} &\coloneqq \left(\omega ^{-1}\bm{\Delta}\right)^{1/2} \ket{u} \eqqcolon \left(\textcolor{black}{ \begin{array}{c}
\textbf{x}_\star \\ \textbf{y}_\star
\end{array}}\right).
\end{align*}
%
Those eigenvectors are such that $\braket{u_\star | u_\star} = \omega ^{-1}  \braket{ u | \bm{\Delta} | u }$ — that is, according to the fact that $\omega ^{-1}  \braket{ u | \bm{\Delta} | u } = \braket{ u |  \textbf{J} | u }$,
\begin{align*}
\braket{u_\star | u_\star} =  \textbf{x}\tvec\textbf{x}-\textbf{y}\tvec\textbf{y},
\end{align*}
i.e., if the condition in \eqref{eq:x2-y2} is fulfilled,
\begin{equation}\label{eq:starnormalized}
\textbf{x}_\star^\top\mkern-2mu\textbf{x}_\star + \textbf{y}_\star^\top\mkern-2mu\textbf{y}_\star = 1.
\end{equation}
The $\textbf{x}_\star$ and $\textbf{y}_\star$ vectors are related to $\textbf{z}_+$ and $\textbf{z}_-$ through $(2\textbf{x}_\star = \textbf{z}_- + \textbf{z}_+)$ and $(2\textbf{y}_\star = \textbf{z}_- - \textbf{z}_+)$.
From these results, Luzanov and Zhikol have suggested to build two normalized ansatzes, namely
\begin{align*}
\psi _\star^x &= \left(\textbf{x}_\star ^\top\mkern-2mu \textbf{x}_\star^{\textcolor{white}{\dag}} \right)^{-1/2} \sum _{i=1}^N \sum _{a=N+1}^L (\textbf{x}_\star)_{ia} \hat{a}^\dag\hat{i}\psi _0, \\
\psi _\star^y &= \left(\textbf{y}_\star ^\top\mkern-2mu \textbf{y}_\star^{\textcolor{white}{\dag}} \right)^{-1/2} \sum _{i=1}^N \sum _{a=N+1}^L (\textbf{y}_\star)_{ia} \hat{a}^\dag\hat{i}\psi _0.
\end{align*}
These two ansatzes are of $\zeta$--type, and lead — using the same manipulations as in (\ref{eq:endrecastsquare}) —, to two state 1--RDM's which read
\begin{align*}
\bm{\gamma}_\star ^x &= \bm{\gamma}_0 + \left(\textbf{x}_\star ^\top \mkern-2mu\textbf{x}_\star^{\textcolor{white}{\dag}} \right)^{-1}\left[ \left(-\textbf{X}_\star^{\textcolor{white}{\dag}} \textbf{X}_\star ^\top\right) \oplus \left(\textbf{X}_\star ^\top \textbf{X}_\star^{\textcolor{white}{\dag}}\right) \right], \\
\bm{\gamma}_\star ^y &= \bm{\gamma}_0 + \left(\textbf{y}_\star ^\top\mkern-2mu \textbf{y}_\star^{\textcolor{white}{\dag}} \right)^{-1}\left[ \left(-\textbf{Y}_\star^{\textcolor{white}{\dag}} \textbf{Y}_\star ^\top\right) \oplus \left(\textbf{Y}_\star ^\top \textbf{Y}_\star^{\textcolor{white}{\dag}}\right) \right] ,
\end{align*}
where $\bm{\gamma}_0$ is the ground-state 1--RDM. They then use a statistical mixture of the corresponding 1--RDM's to build the total excited-state 1--RDM:
\begin{equation*}
\bm{\gamma}_\star = \left(\textbf{x}_\star ^\top \mkern-2mu \textbf{x}_\star^{\textcolor{white}{\dag}} \right)\bm{\gamma}_\star ^x + \left(\textbf{y}_\star ^\top \mkern-2mu \textbf{y}_\star^{\textcolor{white}{\dag}} \right)\bm{\gamma}_\star ^y.
\end{equation*}
According to what precedes in this paper, we can see that, if $\psi_0$ is a one-determinant state, $(\bm{\gamma}_0,\bm{\gamma}_\star)$ leads to the 1--DDM
\begin{equation*}
\bm{\gamma}^\Delta_\star = \left(-\textbf{X}_\star^{\textcolor{white}{\dag}} \textbf{X}_\star ^\top - \textbf{Y}_\star^{\textcolor{white}{\dag}} \textbf{Y}_\star ^\top\right) \oplus \left(\textbf{X}_\star ^\top \textbf{X}_\star^{\textcolor{white}{\dag}}+\textbf{Y}_\star ^\top \textbf{Y}_\star^{\textcolor{white}{\dag}}\right)
\end{equation*}
and to the detachment/attachment density matrices
\begin{align*}
\bm{\gamma}^d_\star &= \left(\textbf{X}_\star^{\textcolor{white}{\dag}} \textbf{X}_\star ^\top + \textbf{Y}_\star^{\textcolor{white}{\dag}} \textbf{Y}_\star ^\top\right) \oplus \textbf{0}_v,\\
\bm{\gamma}^a_\star &= \textbf{0}_o \oplus \left(\textbf{X}_\star ^\top \textbf{X}_\star^{\textcolor{white}{\dag}}+\textbf{Y}_\star ^\top \textbf{Y}_\star^{\textcolor{white}{\dag}}\right),
\end{align*}
which share the same algebraic properties as the matrices in the EOM picture, though they are derived from AMBW's. The difference with the EOM picture is the existence condition — positive semidefiniteness of the (\textbf{A} + \textbf{B}) and (\textbf{A} $-$ \textbf{B}) matrices — for this picture, and, according to \eqref{eq:starnormalized}, the fact that $
\vartheta _\star$, i.e., the trace of the detachment and attachment 1--RDM's, is equal to unity.
\subsubsection{The AMBW's of Subotnik and co-workers}
\noindent An alternative auxiliary scheme was introduced by Subotnik and co-workers for TDHF and TDDFT \cite{alguire_calculating_2015}. It comes with pairs of ansatzes (denoted hereafter by the Greek letter $\lambda$):
\begin{align}
\forall (m,n) \in S^2, \quad {\psi _{m;n}^\lambda} &\coloneqq \left(\hat{X}_{m} + \hat{X}_m\hat{X}_n\hat{Y}_n\right){\psi _0}, \label{eq:lambdapsi1} \\
{\psi _{n;m}^\lambda} &\coloneqq \left(\hat{X}_{n} + \hat{X}_n\hat{X}_m\hat{Y}_m\right){\psi _0},\label{eq:lambdapsi2}
\end{align}
where ``$m;\!n$'' reads ``$m$ given $n$'', and
\begin{align*}
\forall m \in S,\; \hat{X}_m &= \sum _{i=1}^N\sum_{a=N+1}^L (\textbf{x}_m)_{ia}\, \hat{a}^\dag\hat{i}, \\
\hat{Y}_m &= \sum _{i=1}^N\sum_{a=N+1}^L (\textbf{y}_m)_{ia} \,\hat{a}^\dag\hat{i},
\end{align*}
i.e., $\hat{X}$ and $\hat{Y}$ are both \textit{excitation} operators, but with coefficients taken from the components of $\textbf{x}$ and $\textbf{y}$. The expression for the elements of the 1--TDM corresponding to the $({\psi _0} \longrightarrow {\psi _{m;n}^\lambda})$ transition is:
\begin{equation*}
\forall (m,n)\in S^2,\,\forall (r,s)\in C,\,
(\bm{\gamma}_{0\rightarrow m;n}^{\mathrm{T},\lambda})_{s,r} = \braket{\psi _0 | \hat{r}^\dag\hat{s}|{\psi _{m;n}^\lambda}}.
\end{equation*}
We further develop the expression of one element — say, the ``$(s,r)$'' 1--TDM element:
\begin{align}
\braket{\psi _0 | \hat{r}^\dag\hat{s}|{\psi _{m;n}^\lambda}} &= \braket{\psi _0 | \hat{r}^\dag\hat{s}\hat{X}_m|\psi _0} + \braket{\psi _0 | \hat{r}^\dag\hat{s}\hat{X}_m\hat{X}_n\hat{Y}_n|\psi _0}.\label{eq:lambda1TDM}
\end{align}
The second term in \eqref{eq:lambda1TDM} vanishes, and the 1--TDM element reduces to
\begin{align*}\nonumber
\braket{\psi _0 | \hat{r}^\dag\hat{s}|{\psi _{m;n}^\lambda}} &= \sum _{i=1}^N\sum _{a=N+1}^L \textbf{x}_{ia} \braket{ \psi _0 | \hat{r}^\dag \hat{s}\hat{a}^\dag\hat{i} |\psi _0} \nonumber\\
&= \sum _{i=1}^N\sum _{a=N+1}^L \textbf{x}_{ia} \, \delta _{r,i} \delta_{s,a} \,n_r(1-n_s)
\end{align*}
where $n_r$ (respectively, $n_s$) is the occupation number of $\varphi_r$ (respectively, $\varphi _s$) in the $\psi_0$ state. This leads to
\begin{eqnarray*}
 \left( 
 \begin{array}{cc}
\textbf{0}_{o\textcolor{white}{v}}  &\textbf{0}_{ov}    \\
\textbf{X}_m^\top   &  \textbf{0}_v \\
 \end{array}  \right) =  \bm{\gamma}_{0\rightarrow m;n}^{\mathrm{T},\lambda} \in \mathbb{R}^{L \times L}, 
\end{eqnarray*}
with, for every $(i,a)$ in $I\times A$, $(\textbf{x}_m)_{ia} = (\textbf{X}_m)_{i,a-N}$. We see that $\bm{\gamma}_{0\rightarrow m;n}^{\mathrm{T},\lambda}$ has the exact $\zeta$--type 1--TDM structure, but eludes the de-excitation coefficients. On the other hand, one element — say, the ``$(s,r)$'' once again — of the 1--DDM for the $({\psi _0} \longrightarrow {\psi _{m;n}^\lambda})$ transition reads 
\begin{equation*}
(\bm{\gamma}^{\Delta,\lambda}_{m;n})_{s,r} = \braket{\psi _{m;n}^\lambda | \hat{r}^\dag\hat{s}|{\psi _{m;n}^\lambda}} - \braket{\psi _0 | \hat{r}^\dag\hat{s}|{\psi _0}}.
\end{equation*}
The second term reduces to $(\textbf{I}_o\oplus \textbf{0}_v)_{s,r}$,
but the first term becomes
\begin{align*}
\braket{\psi _{m;n}^\lambda | \hat{r}^\dag\hat{s}|{\psi _{m;n}^\lambda}} &= \left\langle \psi _0 \left| \left(\hat{Y}^\dag _n \hat{X}^\dag _n \hat{X}^\dag _m + \hat{X}^\dag _m \right) \hat{r}^\dag \hat{s}  \left( \hat{X}_{m} + \hat{X}_m\hat{X}_n\hat{Y}_n \right)\right| \psi _0\right\rangle
\end{align*}
and has to be developed into four parts:
\begin{align*}
\braket{\psi _{m;n}^\lambda | \hat{r}^\dag\hat{s}|{\psi _{m;n}^\lambda}} = (\textbf{L}_1^{m;n})_{r,s} + (\textbf{L}_2^{m;n})_{r,s} + (\textbf{L}_3^{m;n})_{r,s} + (\textbf{L}_4^{m;n})_{r,s},
\end{align*}
with
\begin{align*}
(\textbf{L}_1^{m;n})_{r,s} &=  \braket{\psi _0 |  \hat{X}^\dag _m  \hat{r}^\dag \hat{s} \hat{X}_{m}  | \psi _0}, \\
(\textbf{L}_2^{m;n})_{r,s} &= \braket{\psi _0 | \hat{Y}^\dag _n \hat{X}^\dag _n \hat{X}^\dag _m   \hat{r}^\dag \hat{s} \hat{X}_{m} | \psi _0}, \\
(\textbf{L}_3^{m;n})_{r,s} &= \braket{\psi _0 |\hat{X}^\dag _m  \hat{r}^\dag \hat{s} \hat{X}_m\hat{X}_n\hat{Y}_n | \psi _0}, \\
(\textbf{L}_4^{m;n})_{r,s} &= \braket{\psi _0 | \hat{Y}^\dag _n \hat{X}^\dag _n \hat{X}^\dag _m   \hat{r}^\dag \hat{s} \hat{X}_m\hat{X}_n\hat{Y}_n | \psi _0}.
\end{align*}
After developing the $\hat{X}$ and $\hat{Y}$ operators, we see that $\textbf{L}_2^{m;n}$ and $\textbf{L}_3^{m;n}$ are two $L\times L$ zero matrices, and that, if we denote the trace of $(\textbf{X}_m^\top\textbf{X}_m)$ by $\vartheta_m$,
\begin{equation*}
\textbf{L}_1^{m;n} = (-\textbf{X}_m^{\textcolor{white}{\dag}}\textbf{X}_m^\top)\oplus (\textbf{X}_m^\top\textbf{X}_m^{\textcolor{white}{\dag}}) + \vartheta_m (\textbf{I}_o\oplus \textbf{0}_v),
\end{equation*}
so that the $\bm{\gamma}^{\Delta,\lambda}_{m;n}$ matrix is a $\zeta$--like 1--DDM only if both the ``$\vartheta_m$ is equal to unity'' and ``$\textbf{L}_4^{m;n}$ is the $L\times L$ zero matrix'' conditions are fulfilled — indeed, there is still the $\textbf{L}_4^{m;n}$ matrix to add to $\textbf{L}_1^{m;n}$ for getting the full $\bm{\gamma}^{\Delta,\lambda}_{m;n}$. Each of the $L^2$ matrix elements of $\textbf{L}_4^{m;n}$ is a sum of $(N(L-N))^6$ reference-state expectation values of chains of fourteen second quantization operators, each expectation value being decomposed into 252 products of Kronecker's deltas: From Corollary \ref{cor:wick} with $(K=7)$, we see that it is possible to rewrite each expectation value as $(14-1)!! = 135135$ products of 7 Kronecker's deltas. Noticing that our chains of second quantization operators are always an alternation of creation and annihilation operators starting with a creation operator reduces the number of products susceptible of being non-zero to $7! = 5040$ — in fact, $K!$ in the general case of a chain of $2K$ second quantization operators structured as an alternation of creation and annihilation operators. Noticing that $\hat{X}$ and $\hat{Y}$ in \eqref{eq:lambdapsi1} and \eqref{eq:lambdapsi2} are \textit{both} excitation operators, so that our chains are all structured as ``$\hat{Q}_1 \, \hat{r}^\dag\hat{s}\,\hat{Q}_2$'' where $\hat{Q}_1$ (respectively, $\hat{Q}_2$) is an alternating chain — starting by a creation operator — of creation and annihilation operators in which the creation operators are all indexed in $I$ (respectively, $A$) while the annihilation operators are all indexed in $A$ (respectively, $I$), we reduce the number products of Kronecker's deltas to $(3!)^2(1+2\times 3) = 252$ — in fact, $(h!)^2(1+2h)$ with $h= (K-1)/2$ in the general case.

We calculated symbolically the expression of the entries of a generic $\textbf{L}_{4}^{m;n}$ matrix. Our results show that $\textbf{L}_{4}^{m;n}$ is block-diagonal, and structured as $(\textbf{L}_{4,o-}^{m;n} + \textbf{L}_{4,o+}^{m;n})\oplus (\textbf{L}_{4,v-}^{m;n} + \textbf{L}_{4,v+}^{m;n})$. The 252 terms of $\textbf{L}_{4}^{m;n}$ are given in Supplementary Information. The 1--DDM is thus finally composed of 254 terms — one from the ground state, one from $\textbf{L}_1$, and 252 from $\textbf{L}_4$.
\section{Conclusion}
We have unveiled and discussed properties of molecular electronic transitions representations issued by TDDFRT calculations and their post-processing into auxiliary many-body wavefunctions for the electronic excited states. The native EOM formulation of TDDFRT was extensively discussed in terms of natural-orbital representation of the transitions — with the important result that the related detachment/attachment density matrices can be derived without requiring the diagonalization of the 1--DDM —, and such a representation was compared with the CIS-like picture derived by related ansatz. The NTO population-transfer approximate picture for a transition has also been discussed in the EOM--TDDFRT framework, as well as the detachment/attachment and transition-hole/transition-electron pictures, for which a disambiguation has been previously proposed in another publication.

Two alternative types of ansatzes were reported: the one introduced by Luzanov and Zhikol, and the one introduced by Subotnik and co-workers. The structure and properties of the relevant matrices (the 1--DDM and the 1--TDM) were resolved for these two types of ansatzes. The detachment/attachment density matrices corresponding to the AMBW by Luzanov and Zhikol, though being issued from a couple of CIS-like expansions, have their structure identical to the one from the EOM picture. On the other hand, the 1--TDM derived from the auxiliary ansatzes in the frame proposed by Subotnik and co-authors eludes the TDDFRT ``de-excitations'' contributions, and the 1--DDM in that model contains multi-excitation contributions, but is structured as a sum of 254 terms.

We would like to conclude this contribution by mentioning that without some disambiguation relatively to the interpretation of the detachment/attachment and transition-hole/transition-electron representations, certain TDDFRT pictures of an electronic transition would have remained either equivocal and arbitrary, or incomplete. 
\section*{Acknowledgements}
\noindent Drs Felix Plasser, Pierre-François Loos, Anthony Scemama, Emmanuel Fromager, Benjamin Lasorne, Matthieu Saubanère, Nicolas Saby, Mario Barbatti, Miquel Huix-Rottlant, Christophe Raynaud and Evgeny Posenitskiy are gratefully acknowledged for very fruitful discussions on the topic.

\section{Appendix — Eigenvalues ordering for the $\zeta$--type 1--DDM when $N>(L-N)$ and when $N=(L-N)$}

\noindent \textit{Case 2}: $N > (L-N)$ — We define, for any ordering of the $(\lambda_p^2)_{p\in A}^{\textcolor{white}{a}}$ $N$--tuple,
\begin{equation*}
\bm{\Omega} \coloneqq \bm{\lambda}\tmat\bm{\lambda}\in \mathbb{R}_+^{(L-N)\times(L-N)},
\end{equation*}
i.e., $
\mathrm{diag}\left(\Omega _1, \dotsc , \Omega _{L-N}\right) = \mathrm{diag}\left( \lambda _1^2, \dotsc , \lambda_{L-N}^2\right)$. 
We have that $
\bm{\lambda\lambda}\tmat = \bm{\Omega}\oplus \textbf{0}_z$, with $\textbf{0}_z$ being the $(2N-L)\times(2N-L)$ zero matrix. According to
\begin{equation*}
\forall p \in \llbracket L-N+1,  N\rrbracket, \, \textbf{o}_p \in \mathrm{coker}\left(\textbf{T}\right),
\end{equation*} 
where the cokernel of \textbf{T} is defined as
\begin{equation*}
\mathrm{coker}\left(\textbf{T}\right) = \left\lbrace \textbf{w}\in \mathbb{R}^{N\times 1} \, : \, \textbf{T}\tmat\textbf{w} = \textbf{0}_{L-N} \right\rbrace ,
\end{equation*}
with $\textbf{0} _{L-N}$ being the $(L-N) \times 1$ column vector, we define two matrices: the first one is the $\textbf{O}^\uparrow _2 \coloneqq (\textbf{o}_1^\uparrow, \dotsc , \textbf{o}_{L-N}^\uparrow)$ matrix, i.e., $(L-N)$--tuple in which the column vectors are ordered such that our solution to the matrix eigenvalue problem for $\textbf{TT}\tmat$ can be rewritten in a way that makes the eigenvalues of $\textbf{TT}\tmat$ sorted in the increasing order, i.e., $\textbf{TT}\tmat \textbf{O}^\uparrow _2 = \textbf{O}^\uparrow _2 \bm{\Omega}^\uparrow$, where
\begin{equation*}
\bm{\Omega}^\uparrow = \mathrm{diag}\left(\Omega _1^\uparrow , \dotsc , \Omega _{L-N}^\uparrow\right).
\end{equation*}
The second matrix is simply $\textbf{O}_{\mathrm{coker}} \coloneqq \left(\textbf{o}_{L-N+1} , \dotsc , \textbf{o}_N\right)$. We now permute the columns of \textbf{O} such that the column vectors belonging to the cokernel of \textbf{T} appear first, and define a new matrix, $\textbf{O}_{>} \coloneqq ( \textbf{O}_{\mathrm{coker}},\textbf{O}^\uparrow _2)$. We then have that our solution to the matrix eigenvalue problem for $\textbf{TT}\tmat$ can be rewritten in a way that makes the eigenvalues of $\textbf{TT}\tmat$ sorted in the increasing order:
\begin{equation*}
\left(\textbf{TT}\tmat\right)\textbf{O}_{>} = \textbf{O}_{>} \left(\textbf{0}_z \oplus \bm{\Omega}^\uparrow\right).
\end{equation*}
We also define $\textbf{V}^\uparrow _2$ and $\textbf{V}^\downarrow _2$ such that, for every $\omega$ in $\left\lbrace \uparrow,\downarrow\right\rbrace$, we have $\textbf{T}\tmat\textbf{T}\textbf{V}_2^\omega = \textbf{V}_2^\omega \bm{\Omega}^\omega$. We finally define $\textbf{O}_2^\downarrow$ such that
\begin{equation*}
\textbf{TT}\tmat\textbf{O}^\downarrow_2 = \textbf{O}^\downarrow_2\left(\bm{\Omega}^\downarrow \oplus \textbf{0}_z\right).
\end{equation*}
This allows us to write the matrix of eigenvalues of $\bm{\gamma}_\zeta ^\Delta$ sorted in a decreasing order as
\begin{equation*}
\textbf{K}_2^\downarrow =  \bm{\Omega}^\downarrow \oplus \textbf{0}_z \oplus \left(-\bm{\Omega}^\uparrow\right)
\end{equation*}
with the corresponding matrix of eigenvectors:
\begin{eqnarray*}
\textbf{M}^\downarrow _2 \coloneqq  \left(
\begin{array}{cc}
\textbf{0}_{o\textcolor{white}{v}}  &\textbf{O}_{>}     \\
 \textbf{V}^\downarrow_2 &  \textbf{0}_v 
 \end{array}\right).
\end{eqnarray*}
On the other hand, the matrix of eigenvalues of $\bm{\gamma}_\zeta ^\Delta$ sorted in a increasing order reads
\begin{equation*}
\textbf{K}^\uparrow_2 = \left(- \bm{\Omega}^\downarrow\right) \oplus \textbf{0}_z \oplus \bm{\Omega}^\uparrow
\end{equation*}
with the corresponding matrix of eigenvectors:
\begin{equation*}
\textbf{M}_2^\uparrow = \textbf{O}_2^\downarrow \oplus \textbf{V}_2^\uparrow.
\end{equation*}
\textit{Case 3}: $N = (L-N)$ — We have, for any ordering of the $(\lambda_p^2)_{p\in I}^{\textcolor{white}{a}}$ $N$--tuple, that
\begin{equation*}
\bm{\Gamma} \coloneqq \bm{\lambda\lambda}\tmat = \bm{\lambda}\tmat\bm{\lambda},
\end{equation*}
with $\bm{\Gamma} = \mathrm{diag}\left(\Gamma _1,\dotsc , \Gamma _N\right) = \mathrm{diag}\left(\lambda _1^2,\dotsc , \lambda _N^2\right).$ We simultaneously have that, for every $\omega$ in $\left\lbrace \uparrow , \downarrow \right\rbrace$, the matrix eigenvalue problem for $\textbf{TT}\tmat$ can be rewritten with the considered eigenvalue ordering as $\textbf{TT}\tmat \textbf{O}^\omega_3 = \textbf{O}^\omega_3 \bm{\Gamma}^\omega$, and the matrix eigenvalue problem for $\textbf{T}\tmat\textbf{T}$ can be rewritten with the considered eigenvalue ordering as $\textbf{T}\tmat\textbf{T} \textbf{V}^\omega_3 = \textbf{V}^\omega_3 \bm{\Gamma}^\omega$. It is then possible to build the matrix of eigenvalues of $\bm{\gamma}^\Delta_\zeta$ sorted in the increasing order as
\begin{equation*}
\textbf{K}_3^\uparrow = \left(-\bm{\Gamma}^\downarrow\right) \oplus \bm{\Gamma}^\uparrow
\end{equation*}
with the corresponding matrix of eigenvectors being $\textbf{M}_3^\uparrow = \textbf{O}_3^\downarrow \oplus \textbf{V}^\uparrow _3$. We can also write the matrix of eigenvalues of $\bm{\gamma}^\Delta_\zeta$ sorted in the decreasing order as
\begin{equation*}
\textbf{K}_3^\downarrow = \bm{\Gamma}^\downarrow \oplus \left(-\bm{\Gamma}^\uparrow\right)
\end{equation*}
with the corresponding matrix of eigenvectors:
\begin{eqnarray*}
\textbf{M}^\downarrow _3 \coloneqq  \left(
\begin{array}{cc}
\textbf{0}_{o\textcolor{white}{v}}  &\textbf{O}_3^\uparrow     \\
 \textbf{V}^\downarrow_3 &  \textbf{0}_v 
 \end{array}\right).
\end{eqnarray*}

\section{Affiliation of the authors}

Enzo Monino\\
\textit{J. Heyrovsk\'{y} Institute of Physical Chemistry, Academy of Sciences of the Czech \mbox{Republic,}} v.v.i. \textit{Dolej\v{s}kova 3, 18223 Prague 8, Czech Republic}

$\;$

\noindent Jérémy Morere and Thibaud Etienne\footnote{thibaud.etienne@univ-lorraine.fr}\\
{\textit{Université de Lorraine,} CNRS, LPCT, \textit{F-54000 Nancy, France}}
\bibliographystyle{ieeetr}

\begin{thebibliography}{10}

\bibitem{casida_time-dependent_1995}
M.~E. Casida, ``Time-{Dependent} {Density} {Functional} {Response} {Theory} for
  {Molecules},'' in {\em Recent {Advances} in {Density} {Functional}
  {Methods}}, vol.~Volume 1 of {\em Recent {Advances} in {Computational}
  {Chemistry}}, pp.~155--192, WORLD SCIENTIFIC, Nov. 1995.

\bibitem{casida_time-dependent_2009}
M.~E. Casida, ``Time-dependent density-functional theory for molecules and
  molecular solids,'' {\em Journal of Molecular Structure: THEOCHEM}, vol.~914,
  pp.~3--18, Nov. 2009.

\bibitem{dreuw_single-reference_2005}
A.~Dreuw and M.~Head-Gordon, ``Single-{Reference} ab {Initio} {Methods} for the
  {Calculation} of {Excited} {States} of {Large} {Molecules},'' {\em Chemical
  Reviews}, vol.~105, pp.~4009--4037, Nov. 2005.
\newblock Publisher: American Chemical Society.

\bibitem{hirata_configuration_1999}
S.~Hirata, M.~Head-Gordon, and R.~J. Bartlett, ``Configuration interaction
  singles, time-dependent {Hartree}–{Fock}, and time-dependent density
  functional theory for the electronic excited states of extended systems,''
  {\em The Journal of Chemical Physics}, vol.~111, pp.~10774--10786, Dec. 1999.
\newblock Publisher: American Institute of Physics.

\bibitem{ferre_density-functional_2016}
N.~Ferré, M.~Filatov, and M.~Huix-Rotllant, eds., {\em Density-{Functional}
  {Methods} for {Excited} {States}}, vol.~368 of {\em Topics in {Current}
  {Chemistry}}.
\newblock Cham: Springer International Publishing, 2016.

\bibitem{adamo_calculations_2013}
C.~Adamo and D.~Jacquemin, ``The calculations of excited-state properties with
  {Time}-{Dependent} {Density} {Functional} {Theory},'' {\em Chemical Society
  Reviews}, vol.~42, pp.~845--856, Jan. 2013.
\newblock Publisher: The Royal Society of Chemistry.

\bibitem{luzanov_electron_2010}
A.~V. Luzanov and O.~A. Zhikol, ``Electron invariants and excited state
  structural analysis for electronic transitions within {CIS}, {RPA}, and
  {TDDFT} models,'' {\em International Journal of Quantum Chemistry}, vol.~110,
  no.~4, pp.~902--924, 2010.
\newblock \_eprint: https://onlinelibrary.wiley.com/doi/pdf/10.1002/qua.22041.

\bibitem{lipkowitz_excited_2009}
P.~Elliott, F.~Furche, and K.~Burke, ``Excited {States} from {Time}-{Dependent}
  {Density} {Functional} {Theory},'' in {\em Reviews in {Computational}
  {Chemistry}} (K.~B. Lipkowitz and T.~R. Cundari, eds.), pp.~91--165, Hoboken,
  NJ, USA: John Wiley \& Sons, Inc., Jan. 2009.

\bibitem{barbatti_nonadiabatic_2011}
M.~Barbatti, ``Nonadiabatic dynamics with trajectory surface hopping method:
  {Dynamics} with surface hopping,'' {\em Wiley Interdisciplinary Reviews:
  Computational Molecular Science}, vol.~1, pp.~620--633, July 2011.

\bibitem{wang_nactddft_2021}
W.~L. Zikuan~Wang, Chenyu~Wu, ``Nac-tddft: Time-dependent density functional
  theory for nonadiabatic couplings,'' {\em arXiv:2105.10804 [physics]}, May
  2021.
\newblock arXiv: 2105.10804.

\bibitem{hu_nonadiabatic_2007}
C.~Hu, H.~Hirai, and O.~Sugino, ``Nonadiabatic couplings from time-dependent
  density functional theory: {Formulation} in the {Casida} formalism and
  practical scheme within modified linear response,'' {\em The Journal of
  Chemical Physics}, vol.~127, p.~064103, Aug. 2007.

\bibitem{hu2010nonadiabatic}
C.~Hu, O.~Sugino, H.~Hirai, and Y.~Tateyama, ``Nonadiabatic couplings from the
  kohn-sham derivative matrix: Formulation by time-dependent density-functional
  theory and evaluation in the pseudopotential framework,'' {\em Physical
  Review A—Atomic, Molecular, and Optical Physics}, vol.~82, no.~6,
  p.~062508, 2010.

\bibitem{li_first-order_2014}
Z.~Li and W.~Liu, ``First-order nonadiabatic coupling matrix elements between
  excited states: {A} {Lagrangian} formulation at the {CIS}, {RPA}, {TD}-{HF},
  and {TD}-{DFT} levels,'' {\em The Journal of Chemical Physics}, vol.~141,
  p.~014110, July 2014.

\bibitem{li_first_2014}
Z.~Li, B.~Suo, and W.~Liu, ``First order nonadiabatic coupling matrix elements
  between excited states: {Implementation} and application at the {TD}-{DFT}
  and pp-{TDA} levels,'' {\em The Journal of Chemical Physics}, vol.~141,
  p.~244105, Dec. 2014.

\bibitem{ipatov_excited-state_2009}
A.~Ipatov, F.~Cordova, L.~J. Doriol, and M.~E. Casida, ``Excited-state
  spin-contamination in time-dependent density-functional theory for molecules
  with open-shell ground states,'' {\em Journal of Molecular Structure:
  THEOCHEM}, vol.~914, pp.~60--73, Nov. 2009.

\bibitem{mclachlan_time-dependent_1964}
A.~D. McLACHLAN and M.~A. BALL, ``Time-{Dependent} {Hartree}-{Fock} {Theory}
  for {Molecules},'' {\em Reviews of Modern Physics}, vol.~36, pp.~844--855,
  July 1964.
\newblock Publisher: American Physical Society.

\bibitem{mayer_using_2007}
I.~Mayer, ``Using singular value decomposition for a compact presentation and
  improved interpretation of the {CIS} wave functions,'' {\em Chemical Physics
  Letters}, vol.~437, pp.~284--286, Apr. 2007.

\bibitem{maurice_configuration_1995}
D.~Maurice and M.~Head-Gordon, ``Configuration interaction with single
  substitutions for excited states of open-shell molecules,'' {\em
  International Journal of Quantum Chemistry}, vol.~56, pp.~361--370, Feb.
  1995.

\bibitem{foresman_toward_1992-1}
J.~B. Foresman, M.~Head-Gordon, J.~A. Pople, and M.~J. Frisch, ``Toward a
  systematic molecular orbital theory for excited states,'' {\em The Journal of
  Physical Chemistry}, vol.~96, pp.~135--149, Jan. 1992.

\bibitem{surjan_natural_2007}
P.~R. Surján, ``Natural orbitals in {CIS} and singular-value decomposition,''
  {\em Chemical Physics Letters}, vol.~439, pp.~393--394, May 2007.

\bibitem{crespo-otero_recent_2018}
R.~Crespo-Otero and M.~Barbatti, ``Recent {Advances} and {Perspectives} on
  {Nonadiabatic} {Mixed} {Quantum}–{Classical} {Dynamics},'' {\em Chemical
  Reviews}, vol.~118, pp.~7026--7068, Aug. 2018.

\bibitem{plasser_surface_2014}
F.~Plasser, R.~Crespo-Otero, M.~Pederzoli, J.~Pittner, H.~Lischka, and
  M.~Barbatti, ``Surface {Hopping} {Dynamics} with {Correlated}
  {Single}-{Reference} {Methods}: {9H}-{Adenine} as a {Case} {Study},'' {\em
  Journal of Chemical Theory and Computation}, vol.~10, pp.~1395--1405, Apr.
  2014.

\bibitem{tapavicza_trajectory_2007}
E.~Tapavicza, I.~Tavernelli, and U.~Rothlisberger, ``Trajectory {Surface}
  {Hopping} within {Linear} {Response} {Time}-{Dependent}
  {Density}-{Functional} {Theory},'' {\em Physical Review Letters}, vol.~98,
  p.~023001, Jan. 2007.

\bibitem{tavernelli_non-adiabatic_2009}
I.~Tavernelli, E.~Tapavicza, and U.~Rothlisberger, ``Non-adiabatic dynamics
  using time-dependent density functional theory: {Assessing} the coupling
  strengths,'' {\em Journal of Molecular Structure: THEOCHEM}, vol.~914,
  pp.~22--29, Nov. 2009.

\bibitem{tavernelli_nonadiabatic_2009}
I.~Tavernelli, E.~Tapavicza, and U.~Rothlisberger, ``Nonadiabatic coupling
  vectors within linear response time-dependent density functional theory,''
  {\em The Journal of Chemical Physics}, vol.~130, p.~124107, Mar. 2009.

\bibitem{tavernelli_nonadiabatic_2009-1}
I.~Tavernelli, B.~F.~E. Curchod, and U.~Rothlisberger, ``On nonadiabatic
  coupling vectors in time-dependent density functional theory,'' {\em The
  Journal of Chemical Physics}, vol.~131, no.~19, p.~196101, 2009.

\bibitem{tavernelli_nonadiabatic_2010}
I.~Tavernelli, B.~F.~E. Curchod, A.~Laktionov, and U.~Rothlisberger,
  ``Nonadiabatic coupling vectors for excited states within time-dependent
  density functional theory in the {Tamm}–{Dancoff} approximation and
  beyond,'' {\em The Journal of Chemical Physics}, vol.~133, p.~194104, Nov.
  2010.

\bibitem{alguire_calculating_2015}
E.~C. Alguire, Q.~Ou, and J.~E. Subotnik, ``Calculating {Derivative}
  {Couplings} between {Time}-{Dependent} {Hartree}–{Fock} {Excited} {States}
  with {Pseudo}-{Wavefunctions},'' {\em The Journal of Physical Chemistry B},
  vol.~119, pp.~7140--7149, June 2015.

\bibitem{plasser_new_2014}
F.~Plasser, M.~Wormit, and A.~Dreuw, ``New tools for the systematic analysis
  and visualization of electronic excitations. {I}. {Formalism},'' {\em The
  Journal of Chemical Physics}, vol.~141, p.~024106, July 2014.

\bibitem{etienne_towards_2021}
T.~Etienne, ``Natural-orbital representation of molecular electronic
  transitions,'' {\em arXiv:2104.11947 [physics]}, Aug. 2022.
\newblock arXiv: 2104.11947v2.

\bibitem{etienne_comprehensive_2021}
T.~Etienne, ``A comprehensive, self-contained derivation of the one-body
  density matrices from single-reference excited-state calculation methods
  using the equation-of-motion formalism,'' {\em arXiv:1811.08849 [physics]},
  Apr. 2021.
\newblock arXiv: 1811.08849v15.

\bibitem{luzanov_application_1976}
A.~V. Luzanov, A.~A. Sukhorukov, and V.~E. Umanskii, ``Application of
  transition density matrix for analysis of excited states,'' {\em Theoretical
  and Experimental Chemistry}, vol.~10, pp.~354--361, July 1976.

\bibitem{martin_natural_2003}
R.~L. Martin, ``Natural transition orbitals,'' {\em The Journal of Chemical
  Physics}, vol.~118, pp.~4775--4777, Mar. 2003.

\bibitem{head-gordon_analysis_1995}
M.~Head-Gordon, A.~M. Grana, D.~Maurice, and C.~A. White, ``Analysis of
  {Electronic} {Transitions} as the {Difference} of {Electron} {Attachment} and
  {Detachment} {Densities},'' {\em The Journal of Physical Chemistry}, vol.~99,
  pp.~14261--14270, Sept. 1995.

\bibitem{etienne_boundary}
E.~Monino, J.~Morere, and T.~Etienne, ``Boundary values for the charge
  transferred during an electronic transition: insights from matrix analysis,''
  {\em arXiv:2104.13465 [physics]}.

\bibitem{de1981inertia}
E.~M. de~S{\'a}, ``On the inertia of sums of hermitian matrices,'' {\em Linear
  Algebra and its Applications}, vol.~37, pp.~143--159, 1981.

\bibitem{rowe_equations--motion_1968}
D.~J. ROWE, ``Equations-of-{Motion} {Method} and the {Extended} {Shell}
  {Model},'' {\em Reviews of Modern Physics}, vol.~40, pp.~153--166, Jan. 1968.

\bibitem{blase_bethesalpeter_2018}
X.~Blase, I.~Duchemin, and D.~Jacquemin, ``The {Bethe}–{Salpeter} equation in
  chemistry: relations with {TD}-{DFT}, applications and challenges,'' {\em
  Chemical Society Reviews}, vol.~47, pp.~1022--1043, Feb. 2018.

\end{thebibliography}


\end{document}